\documentclass[sigconf,screen]{acmart}

\usepackage{pifont}
\usepackage{multirow}
\usepackage[table]{xcolor} 
\usepackage{colortbl}       
\usepackage[most]{tcolorbox}
\usepackage{threeparttable}
\usepackage{caption}
\usepackage{adjustbox}
\tcbuselibrary{skins}
\usepackage{hyperref}
\usepackage{amsmath}
\usepackage[table]{xcolor}  
\usepackage{booktabs} 
\usepackage[linesnumbered,ruled,vlined]{algorithm2e}
\usepackage{bm} 
\usepackage{standalone}
\usepackage{pgfplots}
\usepackage{subcaption}
\usetikzlibrary{calc}
\usetikzlibrary{plotmarks}
\usetikzlibrary{intersections}
\usepgfplotslibrary{fillbetween}
\usepackage{picture}
\usepackage{microtype}
\usepackage{enumitem}
\setlist{topsep=2pt, itemsep=0pt, parsep=0pt, partopsep=2pt}

\pgfplotsset{
  overlap legend/.style={
    legend image code/.code={%
      \draw[gray!70, thick] (0.31cm,0cm) -- (0.6cm,0cm);
      \draw[mark options={solid, fill=red!30, draw=red!50!black}, mark=diamond*]
        plot coordinates {(0.3cm,0cm)};
      \draw[mark options={solid, fill=teal!30, draw=teal!50!black}, mark=*]
        plot coordinates {(0.6cm,0cm)};
    }
  }
}

\DeclareMathAlphabet\mathbfcal{OMS}{cmsy}{b}{n}

\newtcolorbox{quotebox}{colback=white,breakable,boxrule=0.4pt,colframe=black,fonttitle=\bfseries,top=1pt,bottom=1pt,left=1pt,right=1pt, before skip=5pt, after skip=5pt}

\newtcolorbox{implicationbox}{colback=gray!10,boxrule=0.4pt,colframe=black,fonttitle=\bfseries,top=1pt,bottom=1pt,left=1pt,right=1pt}

\newcommand\mybox[2][]{\tikz[overlay]\node[fill=blue!20,inner sep=1.5pt, anchor=text, draw=black, thick,rectangle,#1] {#2};\phantom{#2}}

\newcommand{\approach}{\texttt{MFTune}}

\makeatletter
\tcbset{
    myhbox/.style 2 args={%
        enhanced, 
        breakable,
        colback=white,
        colframe=blue!30!black,
        attach boxed title to top left={yshift*=-\tcboxedtitleheight}, 
        title={#2},
        boxed title size=title, 
        boxed title style={%
            sharp corners, 
            rounded corners=northwest, 
            colback=tcbcolframe, 
            boxrule=0pt,
            coltitle=white,
            top=0.5mm, bottom=0.5mm, 
        },
        underlay boxed title={%
            \path[fill=tcbcolframe] (title.south west)--(title.south east) 
                to[out=0, in=180] ([xshift=3mm]title.east)--
                (title.center-|frame.east)
                [rounded corners=\kvtcb@arc] |- 
                (frame.north) -| cycle; 
        },
        left=2mm, right=2mm, top=1mm, bottom=1mm, 
        boxsep=1mm, 
        boxrule=0.6pt, 
        #1
    }
}   

\makeatother

\newtcolorbox{myhbox}[2][]{myhbox={#1}{#2}}
\newcounter{RQcount} 
\newcommand{\revisioncolor}{black} 
\newcommand{\revision}[1]{{\color{\revisioncolor}#1}}

{\noindent\begin{minipage}[c]{\linewidth}%
\begin{bclogo}[couleur=gray!30,%
                arrondi=0.1,%
                logo=\bclampe,%
                ombre=true]{\normalsize Key Characteristic} {#1}}%
{\end{bclogo}\end{minipage}\vspace{2mm}}

\setcopyright{cc}
\setcctype{by}
\acmDOI{10.1145/3832783.3834376}
\acmYear{2026}
\copyrightyear{2026}
\acmISBN{979-8-4007-2882-2/2026/10}
\acmConference[ASE '26]{Proceedings of the 41st IEEE/ACM International Conference on Automated Software Engineering}{October 12--16, 2026}{Munich, Germany}
\acmBooktitle{Proceedings of the 41st IEEE/ACM International Conference on Automated Software Engineering (ASE '26), October 12--16, 2026, Munich, Germany}
\acmSubmissionID{ase26main-p988-p}
\received{2026-03-26}
\received[accepted]{2026-06-18}

\begin{document}

\title{Less Is More: Tuning Configurable Systems with Imperfect Fidelity}

\author{Yulong Ye}
\orcid{0009-0007-4856-8577}
\affiliation{
  \institution{IDEAS Lab}
  \city{}
  \state{}
  \country{}
}
\affiliation{
  \institution{University of Birmingham}
  \city{Birmingham}
  \country{United Kingdom}
}
\email{YXY382@student.bham.ac.uk}

\author{Miqing Li}
\orcid{0000-0002-8607-9607}
\affiliation{
  \institution{University of Birmingham}
  \city{Birmingham}
  \country{United Kingdom}
}
\email{m.li.8@bham.ac.uk}

\author{Tao Chen}
\orcid{0000-0001-5025-5472}
\authornote{Corresponding author.}
\affiliation{
  \institution{IDEAS Lab}
  \city{}
  \state{}
  \country{}
}
\affiliation{
  \institution{University of Birmingham}
  \city{Birmingham}
  \country{United Kingdom}
}
\email{t.chen@bham.ac.uk}

\begin{abstract}

Configuration tuning is essential for optimizing the performance of highly configurable systems, e.g., throughput or runtime, under a given environment. Yet, this is a challenging process as there can be many options to tune, and configuration measurement is often highly expensive. In this paper, we demonstrate the phenomenon of ``less can be more'': system configuration tuning can be greatly improved with much superior budget utilization by partially tuning under the imperfect-fidelity---an environment that is similar, but cheaper to measure, compared with the concerned perfect-fidelity of environment under which the system should be tuned. We codify a conceptual framework of fidelity for configurable systems, drawing on which allows us to propose \approach, a tuner that proactively explores in the space of $>10^4$ possible imperfect-fidelity settings to approximate a useful one, which strikes for the wideness of tuning. This creates high-quality seeds for the perfect-fidelity, which in turn ensures the tuning depth. Experiment results against $10$ state-of-the-art tuners, obtained from running diverse real-world systems for $19$ months $24 \times 7$, show that \approach~performs considerably better on $83.33$\% cases with up to $19.34\%$  improvement while achieving hours of budget saving in general.

\end{abstract}






\maketitle

\section{Introduction}
\label{section:introduction}

Modern software systems are often highly configurable, exposing numerous tunable configuration options~\cite{DBLP:conf/icse/LiangHC25,gong2024dividable}. The goal thereof is intuitive: by allowing flexible configuration, the system can achieve greater applicability across diverse domains and cater to varying performance requirements, e.g., runtime and throughput~\cite{chen2015toward,chen2018survey,xiang2025dually}. Yet, excessive configurability comes with its own costs: it has been shown that globally 59\% of the performance issues, where performance requirements were severely violated, are attributed to poorly chosen configurations rather than code~\cite{han2016empirical}. With proper configuration, systems could unlock their full performance potential~\cite{jamshidi2016uncertainty}.

Theoretically, the optimal configuration could be identified by exhaustively profiling the system across all possible configurations under a certain environment, e.g., a setting of anticipated workload or job. This, however, is impractical because measuring even a single configuration can be rather costly, taking considerable time and computational resources (e.g., minutes to hours)~\cite{chen2021multi,chen2024adapting,chen2025accuracy}. The tuning efficiency is further exacerbated by the fact that the configuration space is typically high-dimensional and grows exponentially with the number of configuration options~\cite{10.1145/3803859,DBLP:conf/kbse/XiongC25,chen2018femosaa}. For example, \textsc{MySQL}---a database system---offers dozens of tunable configuration options, making exhaustive profiling infeasible~\cite{DBLP:journals/pvldb/ZhangCLWTLC22,xu2015hey,ma2025faster}.


Existing system tuners often focus on advanced algorithm designs with smart heuristics (e.g, \texttt{BestConfig}~\cite{zhu2017bestconfig}), hoping that the tuner would find the promising configurations soon; or leverage a surrogate model to predict configuration performance (e.g., \texttt{PromiseTune}~\cite{chen2025promisetune}), which can be unreliable at times. The key unaddressed question is: how to efficiently utilize the budget in tuning? In this paper, we take a different perspective to explicitly tackle this challenge: we view tuning a configurable system under different environments as exhibiting a certain degree of exactness to each other, resembling the notion of \textit{fidelity}. Therefore, we hypothesize that proactively exploring good configurations found from an imperfect, cheaper environment under which the system is tuned could be beneficial to tuning it for the perfect, more expensive target environment. For example, configurations found by tuning \textsc{PostgreSQL} under a workload of $100$ requests per second might expedite and benefit the tuning under that with $10,000$ requests per second---a concept we borrowed from the paradigm known as multi-fidelity optimization~\cite{klein2017fast,kandasamy2017multi,hu2019multi}. Noteworthily, while similar in spirit, multi-fidelity optimization differs from prior work on knowledge transfer of environments for configuration tuning~\cite{zhang2021restune,van2017automatic,zhang2023efficient}: the former proactively explores and exploits unknown fidelity settings of environments for tuning, whereas the latter relies on reusing what is available from historically explored and structurally similar environments, limiting their ability to adapt to a completely unforeseen environment. 

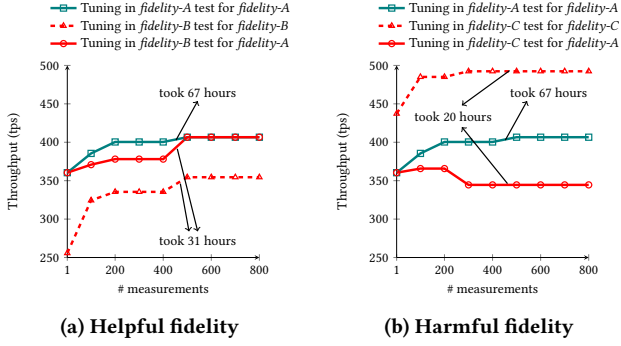
\begin{figure}[!t]
 \centering
  \begin{subfigure}[t]{0.46\columnwidth}
    \centering
\begin{tikzpicture}
    \begin{axis}[
      axis x line  = bottom,
            axis y line  = left  ,
        width=6cm, height=6cm, 
        xlabel={$\#$ measurements},
        ylabel={Throughput (tps)},
        xmin=1, xmax=800,
        xtick={1,200,400,600,800},
        ymin=250, ymax=500,
        ytick={250,300,350,400,450,500},
            legend cell align=left,
          legend columns=1,
        legend style={
          draw=none, fill=none,
          at={(1.2,1.05)}, anchor=south east,
          legend cell align=left
        }
    ]

    \addplot[
        teal,
        mark=square,mark options={thick},
        ultra thick
    ] coordinates {
(1,360.41)
(100,385.49)
(200,400.39)
(300,400.39)
(400,400.39)
(500,406.58)
(600,406.58)
(700,406.58)
(800,406.58)
    }; 

    \addplot[
        red,
        mark=triangle,mark options={thick,solid},
        ultra thick,dashed
    ] coordinates {
(1,255.71)
(100,324.41)
(200,335.43)
(300,335.43)
(400,335.43)
(500,354.49)
(600,354.49)
(700,354.49)
(800,354.49)
    };

        \addplot[
        red,
        mark=o,mark options={thick},
        ultra thick
    ] coordinates {
(1,360.4100)
(100,370.8400)
(200,378.0700)
(300,378.0700)
(400,378.0700)
(500,406.5800)
(600,406.5800)
(700,406.5800)
(800,406.5800)
    };

   
\addlegendentry{Tuning in \textit{fidelity-A} test for \textit{fidelity-A}}
\addlegendentry{Tuning in \textit{fidelity-B} test for \textit{fidelity-B}}
\addlegendentry{Tuning in \textit{fidelity-B} test for \textit{fidelity-A}}
    \end{axis}

    \draw[thick,->] (2.5,2.7) -- (3,3.6);
    \node at (3,3.8) {took $67$ hours};

      \draw[thick,->] (2.53,2.5) -- (3,0.6);
       \draw[thick,->] (2.6,1.7) -- (2.8,0.6);
      \node at (3,0.4) {took $31$ hours};
\end{tikzpicture}
  \subcaption{Helpful fidelity}
  \label{fig:pre-exp1}
  \end{subfigure}
 ~\hspace{0.3cm}
  \begin{subfigure}[t]{0.46\columnwidth}
    \centering
\begin{tikzpicture}
    \begin{axis}[
      axis x line  = bottom,
            axis y line  = left  ,
        width=6cm, height=6cm, 
        xlabel={$\#$ measurements},
        ylabel={Throughput (tps)},
        xmin=1, xmax=800,
        xtick={1,200,400,600,800},
        ymin=250, ymax=500,
        ytick={250,300,350,400,450,500},
            legend cell align=left,
          legend columns=1,
        legend style={
          draw=none, fill=none,
          at={(1.2,1.05)}, anchor=south east,
          legend cell align=left
        }
    ]

    \addplot[
        teal,
        mark=square,mark options={thick},
        ultra thick
    ] coordinates {
(1,360.41)
(100,385.49)
(200,400.39)
(300,400.39)
(400,400.39)
(500,406.58)
(600,406.58)
(700,406.58)
(800,406.58)
    }; 

    \addplot[
        red,
        mark=triangle,mark options={thick,solid},
        ultra thick,dashed
    ] coordinates {
(1,437.29)
(100,485.04)
(200,485.04)
(300,492.38)
(400,492.38)
(500,492.38)
(600,492.38)
(700,492.38)
(800,492.38)
    };

        \addplot[
        red,
        mark=o,mark options={thick},
        ultra thick
    ] coordinates {
(1,360.41)
(100,365.75)
(200,365.75)
(300,344.52)
(400,344.52)
(500,344.52)
(600,344.52)
(700,344.52)
(800,344.52)
    };

   
\addlegendentry{Tuning in \textit{fidelity-A} test for \textit{fidelity-A}}
\addlegendentry{Tuning in \textit{fidelity-C} test for \textit{fidelity-C}}
\addlegendentry{Tuning in \textit{fidelity-C} test for \textit{fidelity-A}}
    \end{axis}

      \draw[thick,->] (2.5,2.7) -- (3,3.6);
    \node at (3.5,3.8) {took $67$ hours};

      \draw[thick,->] (2.6,4.3) -- (1.5,3.5);
       \draw[thick,->] (2.55,1.7) -- (1.5,3);
      \node at (1.3,3.25) {took $20$ hours};
\end{tikzpicture}
    \subcaption{Harmful fidelity}
    \label{fig:pre-exp2}
  \end{subfigure}
  
    \caption{Tuning \textsc{PostgreSQL} under different environments/fidelity settings via a random search tuner. \textit{fidelity-A} (\texttt{time=180s}; \texttt{size=5000k}) is the ``target perfection''; \textit{fidelity-B} (\texttt{time=30s}; \texttt{size=5000k}) and \textit{fidelity-C} (\texttt{time=60s}; \texttt{size=1050k}) are two imperfect fidelity settings. ``Tuning in \textit{fidelity-A} test for \textit{fidelity-B}'' means that the tuning is guided by configurations measured under \textit{fidelity-A} and the best one found therein is progressively tested under \textit{fidelity-B}.}
   \label{fig:pre-exp}
\end{figure}

Figure~\ref{fig:pre-exp} illustrates the idea where the fidelity of environment is set by the limits of execution time and table size in \textsc{Sysbench}~\cite{sysbench}. Suppose our ``target perfection'' is to tune configuration under \textit{fidelity-A} (hence we are only interested in testing therein), Figure~\ref{fig:pre-exp1} shows that tuning under \textit{fidelity-B} produces almost similarly-performing configurations (solid cycle line), when being tested in \textit{fidelity-A}, to those produced by tuning/testing under \textit{fidelity-A} directly (solid square line), but is $\approx36$ hours faster to complete---it might well exceed the result of tuning/testing under \textit{fidelity-A} if given the same wall-clock time. However, this is not straightforward as not all fidelity settings can be useful for \textit{fidelity-A}, especially given the complex dimensions related to environments and fidelity settings for configurable systems: in Figure~\ref{fig:pre-exp2}, even though the time saving remains significant, the tuning guided in \textit{fidelity-C} has severely misled the results when the configurations found therein are tested under \textit{fidelity-A} (solid cycle line vs. solid square line). This is a core issue that has yet been well-addressed in current multi-fidelity optimization.


 

Indeed, multi-fidelity optimization is an established paradigm~\cite{li2018hyperband,falkner2018bohb,awad-ijcai21,klein2017fast,kandasamy2017multi}, particularly for hyperparameter optimization (HPO)~\cite{hutter2011sequential,snoek2012practical,hu2023hydro}. For example, when tuning the hyperparameters for deep neural network training, one can reduce the number of epochs in the training to achieve a low-fidelity measurement, which provides a cheaper yet informative approximation of the final model performance under a perfect, often much higher, epoch setting of fidelity~\cite{li2021mfes}. Yet, despite those advances, adopting this paradigm for general system configuration tuning is not easy, because:





\textbf{1) Mismatched Problem Formulation:} Unlike HPO, system configuration tuning lacks a general multi-fidelity problem formulation that defines how {fidelity} should be modeled and controlled: in HPO, since machine learning models follow similar pipeline, it is well-known that fidelity is related to factors that are highly influential to training time (i.e., the cost in this context), such as training epochs~\cite{domhan2015speeding} and training sample size~\cite{hu2019multi,klein2017fast}. Yet, given the high variety of system domains, there is no known general definition of fidelity for system configuration tuning. 




    
\textbf{2) Incompatible Tuner Assumption:} Most existing multi-fidelity tuners assume a monotonic relationship between fidelity perfection and cost, i.e., more costly measurements always yield a higher exactness approximation to the target perfect environment~\cite{echevarrieta2024speeding,carstensen2025frozen}. While such an assumption is reasonable in HPO, it does not necessarily hold for configurable systems. For instance, training with more epochs generally yields measurements that more closely approximate the performance under the perfect full epochs; in contrast, we found that when tuning the database system \textsc{PostgreSQL}, extending the benchmarking duration does not always yield measurements that better approximate those obtained under a higher, perfect duration runs (see \S\ref{subsection:motivation_challenges}). Moreover, the multi-fidelity tuners in HPO often rely on a restricted single factor to determine fidelity setting (e.g., the epoch)~\cite{li2018hyperband,falkner2018bohb,li2021mfes,hu2019multi}. Such a design overlooks the possibility that more cost-effective fidelity settings can exist in the multi-dimensional fidelity spaces exhibited in configurable systems (see \S\ref{subsection:motivation_challenges}).

To fill the above gap and fully exploit the benefits of fidelity for system configuration tuning, we first present a tailored conceptual framework to codify the definition of fidelity for configurable systems. Drawing on this, we propose \approach, a fidelity-aware configuration tuning tool specifically for systems. What makes \approach~unique is that it proactively exploits/extracts the multi-dimensional fidelity space of more than ten thousands fidelity settings, finding a fair one (i.e., a conceptual ``knee point'' that represents a desirable trade-off between cost and approximation of the perfect-fidelity\footnote{We use the term \textit{perfect-fidelity} to denote the target fidelity setting of environment under which one prefers to tune a configurable system.}) that can accelerate the tuning under the perfect-fidelity setting via a shared archive, hence striking for both the ``wideness'' and ``depth'' of configuration tuning with improved budget utilization.  In a nutshell, our contributions are:





\begin{itemize}
   \item We propose a unified and extensible conceptual framework to define the notions of fidelity for configurable systems, leading to a new problem formulation (\S\ref{section:codifying_fidelity}).
    \item Deriving from the above framework, we present an active fidelity discovery strategy in \approach~to quantify and extract cheaper, yet useful fidelity settings with respect to the perfect-fidelity setting (\S\ref{subsection:EvoFD}).
    
    \item \approach~embeds a mechanism that synergizes tuning between a selected imperfect- and the perfect-fidelity settings with diversity preservation: tuning under the imperfect-fidelity setting, which is less costly yet with a good level of perfection, aims to cover a wide, promising area of the configuration space with good diversity, finding high-quality archived seeds for the perfect-fidelity setting. In contrast, tuning under the perfect-fidelity setting seeks to fully leverage the rich seeds discovered under the imperfect-fidelity setting, encouraging a deep investigation of the tuning directions implied (\S\ref{subsection:LoFEA}--\S\ref{subsection:HiFEA}).

    \item We assess \approach~on six real-world configurable systems against 10 state-of-the-art tuners scaling up to $11,200$ fidelity settings (\S\ref{section:experimental_setup}--\S\ref{section:results_and_analysis}).
\end{itemize}

The results suggest that \approach~considerably outperforms the others, ranking the best on the majority of the systems (5/6) with up to $19.34$\% improvement while generally saving hours of budget. All source code and data are available at: \textcolor{blue}{\texttt{\url{https://github.com/ideas-labo/mftune}}}.

\section{Preliminaries} 
\label{section:preliminaries}


\subsection{System Configuration Tuning}
A configurable system often comprises a set of configuration options, each taking categorical or numerical values. The objective is to identify a configuration that optimizes a specific performance metric (e.g., minimizing runtime or maximizing throughput) of the system under a target environment. This can be formulated as:

\begin{equation}
\label{eq:software_configuration_tuning_problem}
\begin{aligned}
& \arg \min f(\boldsymbol{x}) \text{ 
 or }\arg\max f(\boldsymbol{x}),\\
& \text { s.t. } \sum_{\boldsymbol{x} \in \mathbfcal{X}} \tau(\boldsymbol{x}) \leq \mathcal{B},
\end{aligned}
\end{equation}where $\boldsymbol{x}$ = ($x_1, x_2, \dots, x_d$) is a configuration with the values of $d$ options in configuration space $\boldsymbol{\mathbfcal{X}}$. $f$ is the concerned performance metric. $\tau(\boldsymbol{x})$ denotes the cost of measuring a configuration $\boldsymbol{x}$ under the given environment. $\mathcal{B}$ is the tuning budget, e.g., wall-clock time.

\subsection{Multi-Fidelity Optimization}
Typically, multi-fidelity optimization reduces the cost of expensive measurements by exploiting cheaper approximates of the objective. Here, fidelity is modeled and controlled via a single factor, such as the number of training epochs or the proportion of training data,  while extensions to multiple factors remain rare~\cite{kandasamy2017multi}. Formally, the objective of a solution $\boldsymbol{x} \in \boldsymbol{\mathbfcal{X}}$ measured with budget $\boldsymbol{r}$ can be denoted by $f(\boldsymbol{x}, \boldsymbol{r})$, where larger budgets incur higher cost $c(\boldsymbol{r})$ but are assumed to provide more accurate approximations of the measurements in the perfect-fidelity setting at $\boldsymbol{r}^{*}$. This one-dimensional and cost-oriented view of fidelity underpins most multi-fidelity optimization algorithms in, e.g., HPO~\cite{li2018hyperband, falkner2018bohb, awad-ijcai21, li2021mfes}.



\section{Codifying Fidelity for Configurable Systems}
\label{section:codifying_fidelity}




\subsection{Fidelity Factors}
\label{subsection:fidelity_factors}
Fidelity factors are control variables for the fidelity settings of environments. In HPO~\cite{DBLP:conf/nips/EggenspergerMMF21}, the fidelity factor is commonly related to ``resource types'', such as training epochs, which are general across any models. However, in configuration tuning, the fidelity factors are typically implicit, environment/system-specific, and lack standardized identification criteria. To formalize this notion, in this work, we define a fidelity factor for configurable systems as:


\begin{quotebox}
   \noindent
   \textit{“A variable that modulates the environment under which a configuration is measured but is exogenous to the configurable system.”}
\end{quotebox}

\begin{figure}[t]
    \centering
    \includegraphics[width=\linewidth]{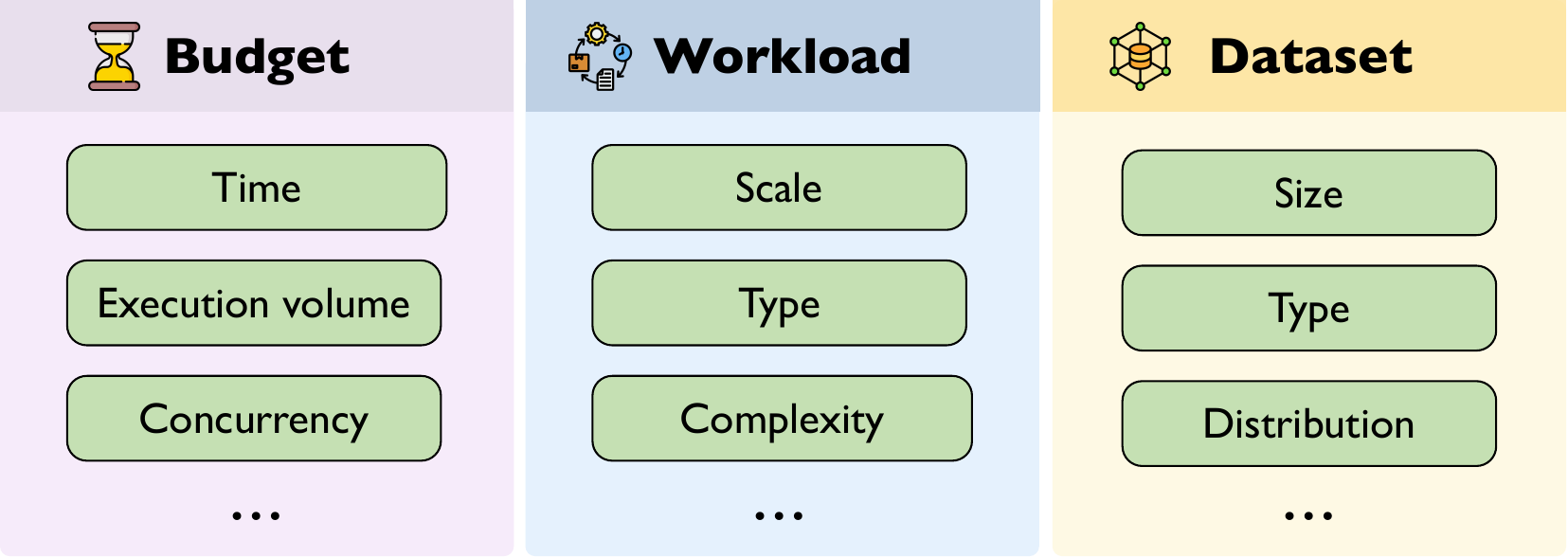}
    \caption{Taxonomy of fidelity factors.}
    \label{fig:taxonomy_fidelity_factor}
\end{figure}

To specify this, we provide a framework of categories grounded in the above definition. Instead of an exhaustive classification, our goal is to offer a structured lens through which potential fidelity factors can be defined, compared, and applied across systems. As illustrated in Figure \ref{fig:taxonomy_fidelity_factor}, the fidelity factors can be assigned into three broad types based on their role and impact on the tuning:



\begin{itemize}
    \item \textbf{Budget-related}: Factors that control resource allocation, e.g., the \texttt{duration}, which controls how long each load-test runs, for tuning \textsc{Tomcat}.
    \item \textbf{Workload-related}: Factors that shape workload characteristics, e.g., the \texttt{max-func} of programs, which sets the maximum number of functions, for tuning \textsc{Gcc}.
    
    \item \textbf{Dataset-related}: Factors that define the dataset composition, e.g., \texttt{table-size} in the benchmark that tunes \textsc{MySQL}.
   
\end{itemize}

In general, a fidelity factor can be categorical, numerical or ordinal, depending on the system/environment. Those fidelity factors can often be found in the benchmark used to tune a configurable system, e.g., \textsc{Sysbench}~\cite{sysbench} and \textsc{Wrk}~\cite{Wrk}.

\subsection{Problem Formulation}

Drawing on the definition of fidelity factors for configurable systems, we extend the problem from Equation~\ref{eq:software_configuration_tuning_problem} as a fidelity-aware one. Suppose $\boldsymbol{z} \in \boldsymbol{\mathbfcal{Z}}$ denotes a fidelity setting, where $\boldsymbol{z}$ = ($z_1, z_2, \dots, z_k$) is a vector defined by $k$ ($k \geq 1$) fidelity factor(s), and $\boldsymbol{\mathbfcal{Z}}$ stands for the entire fidelity space of settings.
The objective function under the fidelity setting $\boldsymbol{z}$ is $f_{\boldsymbol{z}}(\boldsymbol{x})$, i.e., the performance of configuration $\boldsymbol{x}$ measured at $\boldsymbol{z}$. We denote the target, perfect-fidelity setting by $\boldsymbol{z}^*$ ($\boldsymbol{z}^* \in \boldsymbol{\mathbfcal{Z}}$), and its objective function by $f_{\boldsymbol{z}^*}(\boldsymbol{x})$ (i.e., $f_{\boldsymbol{z}^*}(\boldsymbol{x}) = f(\boldsymbol{x})$). We denote the cost under the fidelity setting $\boldsymbol{z}$ by $\tau_{\boldsymbol{z}}(\boldsymbol{x})$. Now, the multi-fidelity configuration tuning problem becomes:




\begin{equation}
\begin{aligned}
& \arg \min f_{\boldsymbol{z}^*}(\boldsymbol{x}) \text{ 
 or }\arg\max f_{\boldsymbol{z}^*}(\boldsymbol{x}),\\
& \text { s.t. } \sum_{\boldsymbol{x} \in \mathbfcal{X}; \text{ } \boldsymbol{z} \in \mathbfcal{Z}} \tau_{\boldsymbol{z}}(\boldsymbol{x}) \leq \mathcal{B}.
\end{aligned}
\end{equation}

The goal is to find a configuration whose performance value under the perfect-fidelity setting is optimized, through exploring the measurements across the entire fidelity space, including both imperfect-fidelity settings ($\mathbfcal{Z}/\boldsymbol{z}^*$) and the perfect-fidelity setting ($\boldsymbol{z}^*$), subject to a budget $\mathcal{B}$. 



From the above, the $i$th fidelity setting $\boldsymbol{z}_i$ for configurable systems might influence two aspects:



\begin{itemize}
    \item \textbf{Cost:} {The cost of an (imperfect-)fidelity setting refers to the average wall-clock time consumed for profiling the system on $u$ given configurations under that setting to obtain their performance measurements\footnote{The time taken for the deployment of the environment is also included.}:}

 \begin{equation}
    \tau_{\boldsymbol{z}_i}=\frac{1}{u} \sum_{j=1}^u \tau_{\boldsymbol{z}_i}(\boldsymbol{x}_j).
 \label{eq:cost}
     \end{equation}
    \item \textbf{Fidelity Perfection:} We use fidelity perfection to denote the extent to which an (imperfect-)fidelity setting can approximate the target perfect-fidelity setting, under which the configurable system should be tuned. To quantify such, in this work we use Spearman correlation~\cite{hauke2011comparison}, which captures the relative nonlinear ranking of the configurations' performance measured under the $i$th (imperfect-)fidelity setting ($\boldsymbol{z}_i$) and the perfect-fidelity setting ($\boldsymbol{z}^*$):

    \begin{equation}
    \rho_{\boldsymbol{z}_i} = \frac{\operatorname{cov}\big(\mathbfcal{R}(\mathbfcal{F}_{\boldsymbol{z}_i}), \mathbfcal{R}(\mathbfcal{F}_{\boldsymbol{z}^*})\big)}
           {\sigma_{\mathbfcal{R}(\mathbfcal{F}_{\boldsymbol{z}_i})} \, \sigma_{\mathbfcal{R}(\mathbfcal{F}_{\boldsymbol{z}^*})}},
    \label{eq:spearman}
    \end{equation}
    whereby $\boldsymbol{\mathbfcal{S}}$ = $\{\boldsymbol{x}_1, \dots, \boldsymbol{x}_u\}$ means a set of $u$ configurations. $\mathbfcal{F}_{\boldsymbol{z}_i}$ = [$f_{\boldsymbol{z}_i}(\boldsymbol{x}_1)$, \dots, $f_{\boldsymbol{z}_i}(\boldsymbol{x}_u)$] and $\mathbfcal{F}_{\boldsymbol{z}^*}$ = [$f_{\boldsymbol{z}^*}(\boldsymbol{x}_1)$, \dots, $f_{\boldsymbol{z}^*}(\boldsymbol{x}_u)$] denote the performance of all configurations $\boldsymbol{x} \in$ $\mathbfcal{S}$ measured under the fidelity setting $\boldsymbol{z}_i$ and the perfect-fidelity setting $\boldsymbol{z}^*$, respectively.  $\mathbfcal{R}(\cdot)$ returns performance ranks; $\operatorname{cov}(\cdot,\cdot)$ is their covariance while
    $\sigma_{\mathbfcal{R}(\mathbfcal{F}_{\boldsymbol{z}_i})}$ and $\sigma_{\mathbfcal{R}(\mathbfcal{F}_{\boldsymbol{z}^*})}$ are the corresponding standard deviations. A higher $\rho_{\boldsymbol{z}_i}$ means the setting is closer to the perfect-fidelity setting.
    
    


\end{itemize}

With the cost and fidelity perfection, we anticipate that most of the time, if not all, the tuning is guided by configurations measured under a selected imperfect-fidelity setting $f_{\boldsymbol{z}_i}(\boldsymbol{x})$, which should often be cheaper yet provides a good perfection.
\subsection{Characteristics of Systems Fidelity}
\label{subsection:motivation_challenges}

To better showcase the characteristics of fidelity in configurable systems, we conduct an exploratory study on \textsc{PostgreSQL} using the \textsc{Sysbench}~\cite{sysbench} benchmark as an example. We tune 20 widely used configuration options and examine 5 commonly mentioned fidelity factors, leading to a configuration space of $8.70 \times 10^{122}$ and fidelity space of 11,200, respectively\footnote{The details can be found at Tables~\ref{tb:systems} and~\ref{tb:workloads}.}. Here, we aim to tune for better throughput while reducing the time taken for profiling configurations (the cost). Specifically, we generate 1,000 configurations using Latin Hypercube Sampling (LHS)~\cite{mckay1992latin} and measure each under a given perfect-fidelity setting and 10 randomly selected imperfect-fidelity settings. We obtain several observations:



\begin{quotebox}
   \noindent
   \textbf{Observation 1:} Top-performing configurations in the imperfect- and perfect-fidelity settings are never completely identical; they can, however, be quite similar or discrepant.
\end{quotebox}

\begin{figure}[t!]
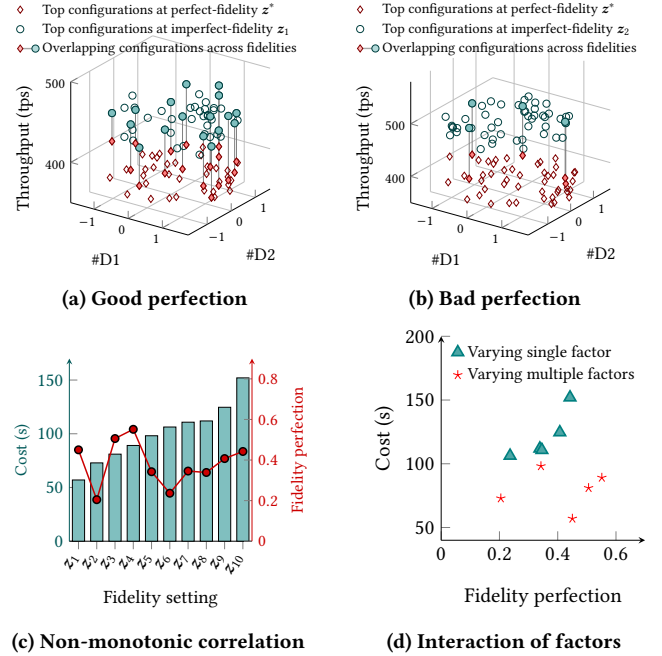

\captionsetup{skip=6pt}
    \centering
    \subfloat[Good perfection\label{fig:ma}]{
        \includegraphics[width=0.46\columnwidth]{figures/motivation_a_new.tex}}
    \hfill
    \subfloat[Bad perfection\label{fig:mb}]{
        \includegraphics[width=0.46\columnwidth]{figures/motivation_b_new.tex}}
        
    \vspace{3pt} 
    
    \subfloat[Non-monotonic correlation\label{fig:mc}]{\begingroup
\pgfplotsset{compat=1.18}

\begin{tikzpicture}

\begin{axis}[
    tick label style={font=\large},
    label style={font=\large},
    legend style={font=\large},
    name=leftaxis,
    height=5cm, width=5cm,
    ybar=0pt,
    bar width=6.5pt,
    xmin=0.5, xmax=10.5,
    ymin=0, ymax=170,
    xlabel={Fidelity setting},
    xtick={1,2,3,4,5,6,7,8,9,10},
    xticklabels={$\boldsymbol{z}_1$,$\boldsymbol{z}_2$,$\boldsymbol{z}_3$,$\boldsymbol{z}_4$,$\boldsymbol{z}_5$,$\boldsymbol{z}_6$,$\boldsymbol{z}_7$,$\boldsymbol{z}_8$,$\boldsymbol{z}_9$,$\boldsymbol{z}_{10}$},
    xticklabel style={rotate=60, anchor=east},
    ylabel={},
    axis y line*=left,
    axis x line*=bottom,
    y axis line style={-stealth,teal!70!black},
    yticklabel style={teal!70!black},
    legend cell align={left},
    legend style={draw=none, fill=none, at={(0.02,0.98)}, anchor=north west},
]
\addplot[fill=teal!50, draw=black] coordinates {
    (1,56.9852)
    (2,72.8931)
    (3,81.0256)
    (4,89.1124)
    (5,98.1484)
    (6,106.2855)
    (7,110.7742)
    (8,111.9248)
    (9,124.6539)
    (10,152.0097)
};
\end{axis}

\begin{axis}[
    name=rightaxis,
    height=5cm, width=5cm,
    at={(leftaxis.south west)}, anchor=south west,
    xmin=0.5, xmax=10.5,
    ymin=0, ymax=0.9,
    axis y line*=right,
    axis x line=none,
    y axis line style={-stealth,red!80!black},
    yticklabel style={red!80!black},
    ylabel={},
    legend cell align={left},
    legend style={draw=none, fill=none, at={(0.98,0.98)}, anchor=north east},
]
\addplot+[
    mark=*,
    thick,
    draw=red!80!black,
    mark options={fill=red!80!black,draw=black}
] coordinates {
    (1,0.4503)
    (2,0.2047)
    (3,0.5059)
    (4,0.5514)
    (5,0.3424)
    (6,0.2365)
    (7,0.3457)
    (8,0.3387)
    (9,0.4072)
    (10,0.4422)
};
\end{axis}

\node[rotate=90, text=teal!70!black, anchor=center]
    at ($(leftaxis.west)+(-25pt,0)$) {Cost (s)};
\node[rotate=90, text=red!80!black, anchor=center]
    at ($(rightaxis.east)+(25pt,0)$) {Fidelity perfection};

\end{tikzpicture}

\endgroup}
    \hfill
    \subfloat[Interaction of factors\label{fig:md}]{\begingroup
\pgfplotsset{compat=1.18}

\begin{tikzpicture}
\begin{axis}[
    width=5cm,
    height=5cm,
    xmin=0, xmax=0.7,
    ymin=40,  ymax=200,
    xlabel={Fidelity perfection},
    ylabel={Cost (s)},
    xlabel style={
    font=\large,
    at={(axis description cs:0.5,-0.2)},  
    anchor=north
    },
    ylabel style={
        font=\large,
        at={(axis description cs:-0.20,0.5)},  
        anchor=south
    },
    legend style={
        font=\small,
         cells={align=left},
        draw=none,            
        fill=none,
        at={(0.02,1)},      
        anchor=north west
    },
    xlabel style={font=\large},
    ylabel style={font=\large},
    tick label style={font=\large},
    axis x line*=bottom,
    axis y line*=left,
    x axis line style={-stealth},
    y axis line style={-stealth},
    enlargelimits=false,
    legend cell align={left}
]

\addplot[
    only marks,
    mark=triangle*,
    mark size=3.2pt,
    line width=0.8pt,
    draw=teal,
    fill=teal!70,
]
table[row sep=\\]{
0.236458 106.285525\\
0.338729 111.924842\\
0.345708 110.774182\\
0.407204 124.653888\\
0.442226 152.009694\\
};
\addlegendentry{Varying single factor};

\addplot[
    only marks,
    mark=star,
    mark size=2.3pt,
    draw=red,
    fill=red,
]
table[row sep=\\]{
0.450334 56.985192\\
0.342429 98.148400\\
0.204739 72.893074\\
0.505940 81.025645\\
0.551390 89.112436\\
};
\addlegendentry{Varying multiple factors};

\end{axis}
\end{tikzpicture}

\endgroup}
    
    \caption{(a) and (b) visualize the landscapes (processed by MDS~\cite{cox2000multidimensional}) of top-50 performing configurations under perfect-fidelity and two distinct imperfect-fidelity settings; (c) presents the cost and fidelity perfection of 10 imperfect-fidelity settings; and (d) shows the selected imperfect-fidelity settings with and without considering factor interactions.}
    \label{fig:motivation}
    \vspace{-10pt}
\end{figure}

Figures~\ref{fig:ma} and~\ref{fig:mb} visualize the top-50 high-performing configurations under the perfect-fidelity setting, which requires about $250$s per measurement, together with those under two representative imperfect-fidelity settings, requiring roughly $60$s and $75$s per measurement, respectively. We see that the top-performing configurations measured under different fidelity settings can rarely be identical, but the level of perfection differs: as shown in Figure~\ref{fig:ma}, there is a strong overlap of the promising configurations between imperfect- and perfect-fidelity settings (the connected points), i.e., the configurations that perform very well under both fidelity settings are identical, evidencing the feasibility of leveraging imperfect-fidelity proxies to guide efficient tuning. However, as illustrated in Figure~\ref{fig:mb}, an opposite pattern emerges, where the top-performing configurations under imperfect- and perfect-fidelity settings differ significantly. This highlights the importance of discovering reliable imperfect-fidelity settings. 


    


As mentioned, HPO assumes a monotonic relation between fidelity perfection and cost (more cost means better perfection), but what we found for configurable systems is that:

\begin{quotebox}
   \noindent
   \textbf{Observation 2:} Cost and fidelity perfection are not always monotonically correlated, making cost variation an unreliable/uncertain indicator of the change to perfection level.
\end{quotebox}

Figure~\ref{fig:mc} plots the fidelity perfection (via Equation~\ref{eq:spearman}) of each imperfect-fidelity setting against its measurement cost. The results show that cost and fidelity perfection are not always monotonically related, e.g., some highly costly settings yield a lower perfection than the cheaper ones. This challenges the common assumption in many multi-fidelity optimization cases that higher cost inherently brings better exactness to the perfect-fidelity~\cite{li2018hyperband,falkner2018bohb,li2021mfes,hu2019multi,carstensen2025frozen}.


To further understand the unique complexity of fidelity space for configurable systems beyond what was assumed for HPO, within the 10 imperfect-fidelity settings, we randomly choose five that differ only on a single factor, e.g., (\texttt{time=180s}) vs. (\texttt{time=30s}), against the other five that differ on multiple factors, e.g., (\texttt{time=180s; tables=50}) vs. (\texttt{time=30s;tables=20}), based on all of which we measure the performance of $1,000$ configurations. We found that:
\begin{quotebox}
   \noindent
   \textbf{Observation 3:} Considering interaction between multiple fidelity factors can yield ``fair'' imperfect-fidelity settings\footnote{Not perfect, but fair enough to be useful given its cost/perfection level.} that would otherwise be difficult to find.
\end{quotebox}

As in Figure~\ref{fig:md}, the imperfect-fidelity settings where multiple factors are varied reflect much better cost and fidelity perfection than their single-factor varying counterparts. This implies that assuming imperfect-fidelity settings with only a single factor, as in the classic HPO, would leave their full potential untapped.

The above, together with the proposed conceptual framework of fidelity, motivates our idea on proactively exploiting imperfect-fidelity settings to accelerate tuning under the perfect-fidelity setting. Yet, this raises three key challenges: 



\begin{itemize}
    \item \textbf{Challenge 1}: How can we identify the fair imperfect-fidelity settings in multi-dimensional fidelity space?
    \item \textbf{Challenge 2}: How to explore useful information under the imperfect-fidelity setting(s)?
    \item \textbf{Challenge 3}: How to enable effective exploitation of the imperfect-fidelity setting to benefit tuning under the perfect-fidelity setting?
\end{itemize}


Those are what we address via \approach.


\section{The \approach~Framework}
\label{section:methodology}





\begin{figure}[t]
    \centering
    \includegraphics[width=\linewidth]{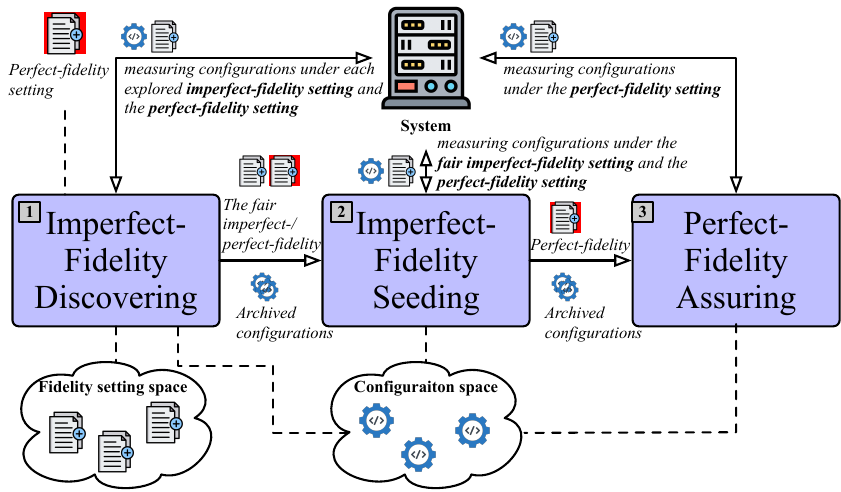}
    \caption{Workflow and architecture overview of \approach.}
    \label{fig:MFTune}
\end{figure}

Figure~\ref{fig:MFTune} and Algorithm~\ref{alg:mftune} illustrate the workflow of \approach. Given a perfect-fidelity setting, the key idea is to first explore the fidelity space to identify a ``fair'' imperfect-fidelity setting---one that offers good fidelity perfection while incurring a much lower cost, hence improving budget utilization. We then conduct a specifically designed diversity-preserving tuning under the selected fair imperfect-fidelity setting to achieve ``wide'' tuning, after which the resulting archive seeds the subsequent ``deep'' tuning under the perfect-fidelity setting. To that end, \approach~comprises three phases:

    \textbf{Imperfect-Fidelity Discovering (line 1)} formulates an active, multi-objective fidelity discovery problem (maximizing level of perfection while minimizing cost) and applies NSGA-II~\cite{deb2002fast}---a common multi-objective genetic algorithm---to evolve a ``fair'' imperfect-fidelity setting, together with some initially archived seeds of configurations (\textit{\textbf{Challenge~1}}).

     \textbf{Imperfect-Fidelity Seeding (line 2)} embeds the discovered imperfect-fidelity setting into the tuning, together with a two-stage tuning strategy, which covers a wide area of the space and generates high-quality configurations to improve archived seeds for tuning under the perfect-fidelity setting (\textit{\textbf{Challenges 2}} and \textit{\textbf{3}}).

     \textbf{Perfect-Fidelity Assuring (line 3)} initializes the tuning by filtering the archive of seeded configurations, hence fully exploiting the benefits from the tuning thereof for deeply exploring along the tuning directions implied and returns the best configuration found under the perfect-fidelity setting (\textit{\textbf{Challenge 3}}).


We equally split the budget into the above: tuning under the fair imperfect-fidelity setting for seeds has half the budget (since it has two-stage tuning), while the other two share a quarter of the budget each. All phases are connected by an archive $\mathbfcal{A}$, which contains configurations progressively measured under the perfect-fidelity setting; hence, the tuning can stop whenever needed. 



\begin{algorithm}[t]
\caption{\approach}
\label{alg:mftune}
\SetAlgoLined
\scriptsize

\KwIn{Budget $\mathcal{B}$; perfect-fidelity setting $\boldsymbol{z}^*$; parameter $\alpha$}


\KwOut{The best configuration found under perfect-fidelity: $\boldsymbol{x}_{best}$}


$\boldsymbol{\mathbfcal{A}}$ , $\boldsymbol{z}_{fair} \gets $\textsc{ImperfectFidelityDiscovering}($\mathcal{B}$, $\boldsymbol{z}^*$)\\


$\boldsymbol{\mathbfcal{A}}$ $\gets$ \textsc{ImperfectFidelitySeeding}($\mathcal{B}$, $\boldsymbol{z}_{fair}$, $\boldsymbol{z}^*$, $\boldsymbol{\mathbfcal{A}}$, $\alpha$)\\

$\boldsymbol{x}_{best} \gets$ \textsc{SeededPerfectFidelityTuning}($\mathcal{B}$, $\boldsymbol{\mathbfcal{A}}$, $\boldsymbol{z}^*$)
   
\Return $\boldsymbol{x}_{best}$
\end{algorithm}

\subsection{Discovering Fair Imperfect-Fidelity}
\label{subsection:EvoFD}

\begin{figure}[t]
    \centering
    \includegraphics[width=\linewidth]{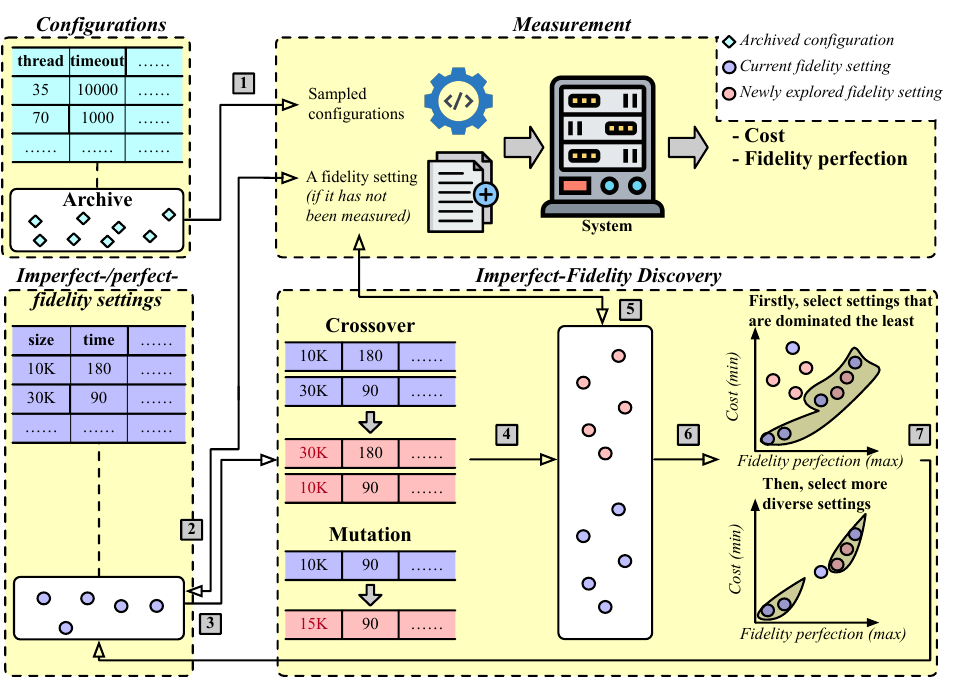}
    \caption{Discovering the fair imperfect-fidelity setting.}
    \label{fig:part1}
\end{figure}

At the beginning, \approach~aims to identify a fair imperfect-fidelity that has lower cost while exhibiting a good level of fidelity perfection to the target perfect-fidelity. As such, this forms a typical multi-objective search problem, for which we adopt the NSGA-II~\cite{deb2002fast} algorithm to explore the interactions between fidelity factors, together with their nonlinear implications to cost and fidelity perfection (\textit{\textbf{Observations 2}} and \textit{\textbf{3}}). Specifically, it runs in the following step with a quarter budget (also shown in Figure~\ref{fig:part1} and Algorithm~\ref{alg:EvoFD}):

\begin{enumerate}
    \item Generate an archive $\boldsymbol{\mathbfcal{A}}$ = $\{\boldsymbol{x}_1, \dots, \boldsymbol{x}_l\}$ with $l$ diverse configurations via Latin Hypercube Sampling (LHS)~\cite{mckay1992latin} to provide the basis for fidelity perfection (~\mybox[fill=gray!30]{1}~).
    \item Initialize a fidelity population $\boldsymbol{\mathbfcal{Q}}$ = \{$\boldsymbol{z}_1, \dots, \boldsymbol{z}_m$\} by randomly sampling $m$ imperfect-fidelity settings (~\mybox[fill=gray!30]{2}~). 
    \item Measure the configurations in $\boldsymbol{\mathbfcal{A}}$ under every fidelity setting in $\boldsymbol{\mathbfcal{Q}}$ and the perfect-fidelity setting $\boldsymbol{z}^*$. These allow us to measure
    the cost (using Equation~\ref{eq:cost}) and fidelity perfection (using Equation~\ref{eq:spearman}) for each of the imperfect-fidelity settings in $\boldsymbol{\mathbfcal{Q}}$ (~\mybox[fill=gray!30]{1}~--~\mybox[fill=gray!30]{2}~).
    \item Reproduce $m$ new imperfect-fidelity settings based on $\boldsymbol{\mathbfcal{Q}}$ via boundary mutation and uniformed crossover operators\footnote{Boundary mutation perturbs a decision variable by resampling it within its feasible range, while uniform crossover exchanges variables between two parent solutions.} in NSGA-II and measure the cost and fidelity perfection under newly explored imperfect-fidelity settings, if any, using the configurations in $\boldsymbol{\mathbfcal{A}}$ (~\mybox[fill=gray!30]{3}~--~\mybox[fill=gray!30]{5}~).
    \item Keep the top $m$ imperfect-fidelity settings among the old and new ones as the new $\boldsymbol{\mathbfcal{Q}}$ via the non-dominated sorting\footnote{Non-dominated sorting ranks solutions into Pareto fronts based on dominance relations, where a solution is non-dominated if no other solution is better in all objectives, followed by crowding distance~\cite{deb2002fast}.} in NSGA-II on both the cost and fidelity perfection (~\mybox[fill=gray!30]{6}~).
    \item Repeat from Step 4 when there are still budgets left; otherwise, return the fair imperfect-fidelity setting with the biggest $\frac{\rho_{z_i}}{\tau_{z_i}}$ from the final population $\boldsymbol{\mathbfcal{Q}}$ (~\mybox[fill=gray!30]{7}~).
\end{enumerate}

The produced archive $\boldsymbol{\mathbfcal{A}}$ and the fair imperfect-fidelity setting, denoted as $\boldsymbol{z}_{fair}$, would be used for archiving more high-quality configuration seeds to improve the ``wideness'' of configuration tuning thereafter.


\begin{algorithm}[t]
\caption{\textsc{ImperfectFidelityDiscovering}}
\label{alg:EvoFD}
\SetAlgoLined
\scriptsize
\KwIn{Budget $\mathcal{B}$; perfect-fidelity setting $\boldsymbol{z}^*$; consumed budget $\tau=0$}
\KwOut{The archive $\boldsymbol{\mathbfcal{A}}$ and the fair imperfect-fidelity setting $\boldsymbol{z}_{fair}$}

$\boldsymbol{\mathbfcal{A}}\gets$ generate $l$ configurations  using LHS\\
$\boldsymbol{\mathbfcal{Q}}\gets$ randomly generate $m$ imperfect-fidelity settings and measure their cost and fidelity perfection using $\boldsymbol{\mathbfcal{A}}$ (via Equations \ref{eq:cost} and \ref{eq:spearman})\\
$\tau \gets$ update total time taken\\

\While{$\tau \textless \frac{\mathcal{B}}{4}$}{
    $\boldsymbol{\mathbfcal{Q}'}\gets$ reproduce $m$ new imperfect-fidelity settings from $\boldsymbol{\mathbfcal{Q}}$ with the boundary mutation and uniformed crossover~\cite{chen2021multi}, and measure their cost and fidelity perfection using $\boldsymbol{\mathbfcal{A}}$ (via Equations \ref{eq:cost} and \ref{eq:spearman})\\
    $\tau \gets$ update total time taken\\
    $\boldsymbol{\mathbfcal{Q}}$ $ \gets$ nondominated sorting on $\boldsymbol{\mathbfcal{Q}}$ $\cup$ $\boldsymbol{\mathbfcal{Q}'}$ as in NSGA-II\\
    
}

\Return $\boldsymbol{\mathbfcal{A}}$ , $\boldsymbol{z}_{fair} \gets$ the imperfect-fidelity setting from $\boldsymbol{\mathbfcal{Q}}$ with the biggest $\frac{\rho_{z_i}}{\tau_{z_i}}$
\end{algorithm}

\subsection{Finding Seeds under Imperfect-Fidelity}
\label{subsection:LoFEA}

Here, \approach~explores and archives high-quality configuration seeds under the fair imperfect-fidelity setting $\boldsymbol{z}_{fair}$ (with a low measurement cost). Indeed, directly tuning under $\boldsymbol{z}_{fair}$ might reflect the tuning under the perfect-fidelity setting $\boldsymbol{z}^*$, but at the same time, it might also mislead it into unwanted local optima, since it remains exhibiting ``imperfection'' with respect to $\boldsymbol{z}^*$ (\textit{\textbf{Observation 1}}). As such, to improve the likelihood of archiving more high-quality seeds for the perfect-fidelity tuning later on, we equip a diversity-preserved, ``wideness-oriented'' two-stage tuning under $\boldsymbol{z}_{fair}$ in \approach, where the former stage globally samples for diverse configurations progressively while the latter aims for stable convergence of configurations with local diversity preservation that can benefit the tuning under $\boldsymbol{z}^*$. Each of the above stages owns a quarter of the budget. In this way, \approach~ensures that it archives the best configurations found under $\boldsymbol{z}_{fair}$, together with some diverse configurations that are of sub-optimal performance at $\boldsymbol{z}_{fair}$, but might still perform well under $\boldsymbol{z}^*$. 


\begin{algorithm}[t]
\caption{\textsc{ImperfectFidelitySeeding}}
\label{alg:LoFEA}
\SetAlgoLined
\scriptsize
\KwIn{Budget $\mathcal{B}$; fair imperfect-fidelity setting $\boldsymbol{z}_{fair}$; perfect-fidelity setting $\boldsymbol{z}^*$; the archive $\boldsymbol{\mathbfcal{A}}$; the probability $\alpha$; consumed budget $\tau=0$}
\KwOut{The updated archive $\boldsymbol{\mathbfcal{A}}$} 

\tcc{\textcolor{blue}{Progressive Diversity Sampling}}

$\{\boldsymbol{x}_1, \dots, \boldsymbol{x}_{p}\} \gets$ sample $p=\frac{\mathcal{B}}{4\times\tau_{\boldsymbol{z}_{fair}}}$ configurations via LHS\\
\While{$\tau < \frac{\mathcal{B}}{4}$}{
    \ForEach{\textnormal{$\boldsymbol{x}$} $\in$ $\{\boldsymbol{x}_1, \dots, \boldsymbol{x}_{p}\}$}{
    Measure $\boldsymbol{x}$ under $\boldsymbol{z}_{fair}$ and update $\tau$\\
    \If{\textnormal{$\boldsymbol{x}$} is currently best under $\boldsymbol{z}_{fair}$ \textbf{ and } \textnormal{$\boldsymbol{x}$} $\notin \mathbfcal{A}$ \textbf{ and } $rand() < \alpha$ }{
      Measure $\boldsymbol{x}$ under $\boldsymbol{z}^*$ and update $\tau$\\
    $\boldsymbol{\mathbfcal{A}}$ $\gets$ $\boldsymbol{\mathbfcal{A}}$ $\cup$ \{$\boldsymbol{x}$\} \\
    } 
    }
}
$\boldsymbol{\mathbfcal{P}}$ $\gets$ top $n$ configurations from $\{\boldsymbol{x}_1, \dots, \boldsymbol{x}_{p}\}$ under $\boldsymbol{z}_{fair}$\\

$\boldsymbol{\mathbfcal{A}}\gets\boldsymbol{\mathbfcal{A}}\cup\boldsymbol{\mathbfcal{P}}$ without redundancy\\

\tcc{\textcolor{blue}{Diversity-Preserved Tuning under Imperfect-Fidelity}}

Reset consumed budget $\tau$ = 0\\

Measure any unmeasured configurations in $\boldsymbol{\mathbfcal{A}}$ under $\boldsymbol{z}^*$; return if all the next budget of $\frac{\mathcal{B}}{4}$ is exhausted, otherwise let the consumed budget be $\tau'$\\

$\tau$ = $\tau'$\\

\While{$\tau \textless \frac{\mathcal{B}}{4}$}{

 $\boldsymbol{\mathbfcal{P}}'$ $\gets$ reproduce $n$ new configurations from $\boldsymbol{\mathbfcal{P}}$ with the boundary mutation and uniformed crossover~\cite{chen2021multi}, and measure their performance under $\boldsymbol{z}_{fair}$\\
    $\tau$ $\gets$ update total time taken\\
    $\boldsymbol{\mathbfcal{P}}$ $\gets$ top $n$ performing configurations from $\boldsymbol{\mathbfcal{P}}' \cup \boldsymbol{\mathbfcal{P}}$ under $\boldsymbol{z}_{fair}$\\

$\boldsymbol{x} \gets$ the best configuration from $\boldsymbol{\mathbfcal{P}}$ under $\boldsymbol{z}_{fair}$\\

 \If{\textnormal{$\boldsymbol{x}$} $\notin \mathbfcal{A}$}{
 Measure $\boldsymbol{x}$ under $\boldsymbol{z}^*$ and update $\tau$\\
  $\boldsymbol{\mathbfcal{A}}$ $\gets$ $\boldsymbol{\mathbfcal{A}}$ $\cup$ \{$\boldsymbol{x}$\}\\

  }

}

$\boldsymbol{\mathbfcal{A}}\gets\boldsymbol{\mathbfcal{A}}\cup\boldsymbol{\mathbfcal{P}}$ without redundancy\\

  \Return The archive $\boldsymbol{\mathbfcal{A}}$\\

\end{algorithm}

\subsubsection{Progressive Diversity Sampling}

\begin{figure}[t]
    \centering
    \includegraphics[width=\linewidth]{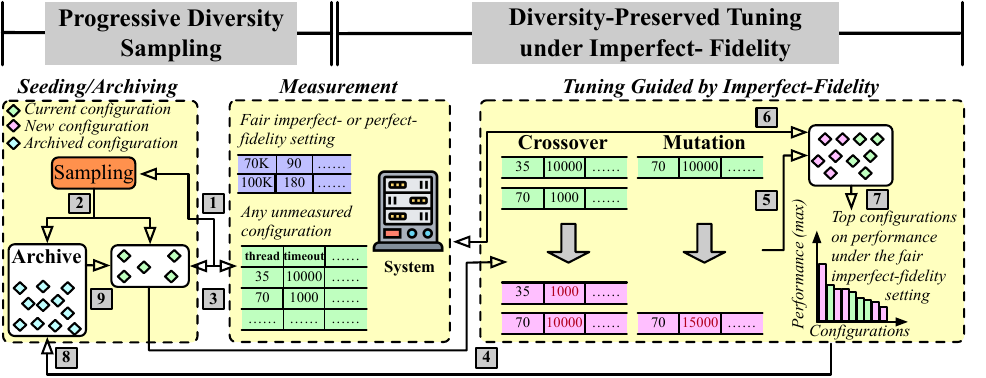}
    \caption{Finding seeds under fair imperfect-fidelity setting.}
    \label{fig:part2}
\end{figure}

To mitigate the imperfection in $\boldsymbol{z}_{fair}$, \approach~adopts a ``top-down'' strategy: globally, an initial set of configurations is measured, filtered, and archived progressively within a quarter budget (~\mybox[fill=gray!30]{1}~--~\mybox[fill=gray!30]{3}~in Figure~\ref{fig:part2} and Algorithm~\ref{alg:LoFEA}): 

\begin{enumerate}
    \item Diversely sample $\frac{\mathcal{B}}{4\times\tau_{\boldsymbol{z}_{fair}}}$ configurations using LHS. In this way, the sampled configurations can be proportional to the budget and cost of $\boldsymbol{z}_{fair}$ (line 1).
    \item Measure a randomly chosen configuration from the above set under $\boldsymbol{z}_{fair}$. If, during the process, a configuration has the best performance for $\boldsymbol{z}_{fair}$ so far, then place it into the archive $\boldsymbol{\mathbfcal{A}}$ with a probability of $\alpha$ (and measure its performance under $\boldsymbol{z}^*$ if that is the case). This controllable parameter, together with the order of configurations, is the key to creating global diversity of preserved configurations in the archive that will seed the tuning under $\boldsymbol{z}^*$ (lines 2-10).
    \item Repeat from 2 until all sampled configurations are measured or the quarter budget is exhausted.
\end{enumerate}

Finally, those top $n$ configurations are ready for the subsequent stage of tuning under $\boldsymbol{z}_{fair}$. To prevent discarding well-performing configurations under $\boldsymbol{z}_{fair}$, we also archive any previously unarchived top $n$ configurations for $\boldsymbol{z}_{fair}$ (over all measured ones).

\subsubsection{Diversity-Preserved Tuning under Imperfect-Fidelity}

As from \mybox[fill=gray!30]{3}~--~\mybox[fill=gray!30]{9} in Figure~\ref{fig:part2} and Algorithm~\ref{alg:LoFEA}, to tune under the fair imperfect-fidelity setting $\boldsymbol{z}_{fair}$, we follow the ``bottom-up'' process in Genetic Algorithm (GA)~\cite{back1993overview}, where a population of configurations are reproduced by boundary mutation and uniformed crossover, after which the top $n$ measured configurations under $\boldsymbol{z}_{fair}$ are preserved into the next iteration. Here, we not only archive those best configurations found under $\boldsymbol{z}_{fair}$, but also those sub-optimal ones explored under $\boldsymbol{z}_{fair}$ during the tuning, hence enforcing local diversity.

The budget for this is again capped as a quarter. Notably, we make several extensions to GA for \approach:

\begin{itemize}
    \item The initial set of configurations are the top $n$ ones obtained (under $\boldsymbol{z}_{fair}$) from the progressive diversity sampling. We also ensure that all configurations in $\boldsymbol{\mathbfcal{A}}$ so far have their performance values under $\boldsymbol{z}^*$ (lines 11-15).
    \item At each iteration, the best configuration measured under $\boldsymbol{z}_{fair}$ so far is also pushed into the archive $\boldsymbol{\mathbfcal{A}}$ and measured under $\boldsymbol{z}^*$ (lines 16-25). This maintains local diversity by tracking the progress of tuning, hence further mitigating the possible ``imperfection'' of $\boldsymbol{z}_{fair}$ to $\boldsymbol{z}^*$. 
    \item At the end, all the persevered but non-archived configurations would be pushed to $\boldsymbol{\mathbfcal{A}}$ and measured under $\boldsymbol{z}^*$ (line 26)---this seeks to align with \textit{\textbf{Observation 1}} that well-performing configurations under different fidelity settings might also be consistent.
\end{itemize}


The above consolidates the archive for more high-quality seeds that are either sub-optimal or optimal as measured under $\boldsymbol{z}_{fair}$, since they could both be beneficial for tuning under $\boldsymbol{z}^*$.

\subsection{Assuring under Perfect-Fidelity with Seeds}
\label{subsection:HiFEA}

\begin{figure}[t]
\captionsetup{skip=8pt}
    \centering
    \includegraphics[width=\linewidth]{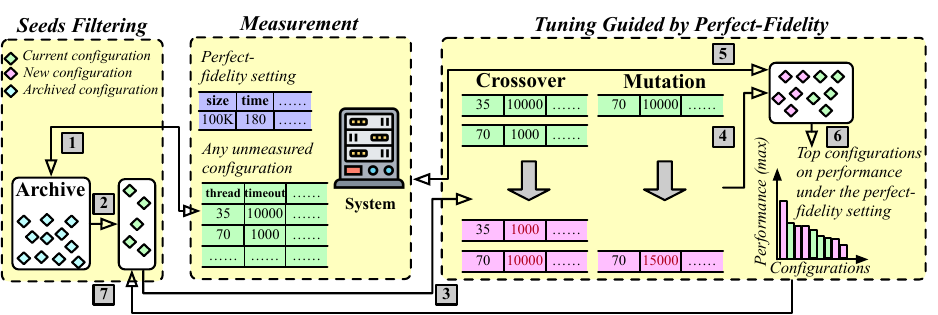}
    \caption{Seeded tuning under perfect-fidelity setting.}
    \label{fig:part3}
  
\end{figure}

Tuning under the perfect-fidelity setting $\boldsymbol{z}^*$ resembles a single-fidelity tuning (Figure~\ref{fig:part3} and Algorithm~\ref{alg:HiFEA}); we again use GA with boundary mutation/uniform crossover (~\mybox[fill=gray!30]{3}~--~\mybox[fill=gray!30]{7}~). Unlike classic GA, the archive $\boldsymbol{\mathbfcal{A}}$ accumulated so far serve as good starting points. \approach~firstly measures all configurations in $\boldsymbol{\mathbfcal{A}}$, which have not been measured, under $\boldsymbol{z}^*$. The configurations in $\boldsymbol{\mathbfcal{A}}$ would be filtered, such that only the top $n$ therein can be used as seeds, enabling a warm start (~\mybox[fill=gray!30]{1}~--~\mybox[fill=gray!30]{2}~). The tuning terminates and returns the best configuration found under $\boldsymbol{z}^*$ when the quarter budget runs out.


This final phase plays a pivotal role for the ``depth'' of tuning in \approach, offering extra assurance to the imperfection/uncertainty of the fair imperfect-fidelity setting.

\begin{algorithm}[t]
\caption{\textsc{SeededPerfectFidelityTuning}}
\label{alg:HiFEA}
\SetAlgoLined
\scriptsize
\KwIn{Budget $\mathcal{B}$;  the archive $\boldsymbol{\mathbfcal{A}}$; perfect-fidelity setting $\boldsymbol{z}^*$; $\tau=0$} 
\KwOut{The best configuration $\boldsymbol{x}_{best}$ found under $\boldsymbol{z}^{*}$} 



Measure any unmeasured configurations in $\boldsymbol{\mathbfcal{A}}$ under $\boldsymbol{z}^*$; return if all the budget of $\frac{\mathcal{B}}{4}$ is exhausted, otherwise let the consumed budget be $\tau'$\\
$\tau=\tau'$\\
$\boldsymbol{\mathbfcal{P}}$ $\gets$ top $n$ configurations from $\boldsymbol{\mathbfcal{A}}$ under $\boldsymbol{z}^{*}$\\

\While{$\tau \textless \frac{\mathcal{B}}{4}$}{

 $\boldsymbol{\mathbfcal{P}}'$ $\gets$ reproduce $n$ new configurations from $\boldsymbol{\mathbfcal{P}}$ with the boundary mutation and uniformed crossover~\cite{chen2021multi}, and measure their performance under $\boldsymbol{z}^{*}$\\
    $\tau$ $\gets$ update total time taken\\
    $\boldsymbol{\mathbfcal{P}}$ $\gets$ top $n$ high-performing configurations from $\boldsymbol{\mathbfcal{P}}' \cup \boldsymbol{\mathbfcal{P}}$ under $\boldsymbol{z}^{*}$\\
}

\Return The best configuration from $\boldsymbol{\mathbfcal{P}}$ measured under $\boldsymbol{z}^{*}$\\

\end{algorithm}

\section{Experimental Setup}
\label{section:experimental_setup}


To evaluate \approach~, we ask four research questions (\textbf{RQs}):

\begin{itemize}
    \item \textbf{RQ1:} How beneficial are the imperfect-fidelity settings?
    \item \textbf{RQ2:} How does \approach~perform compared with the state-of-the-art tuners?
    \item \textbf{RQ3:} How does the imperfect-fidelity discovery and its two-stage tuning individually contribute? 
    \item \textbf{RQ4:} What is the sensitivity of \approach~to $\alpha$?

\end{itemize}

All experiments are independently performed on Ubuntu 22.04.4 LTS running on a cluster of 20 servers, each of which is an Intel NUC with 16 CPU cores and 64GB RAM. In total, the full experiment takes over 19 months of CPU time 24$\times$7.




\subsection{Systems, Options, and Fidelity}

\textbf{Systems:} 
As in Table~\ref{tb:systems}, we select six widely used systems spanning diverse application domains and performance objectives while covering configuration space sizes between $10^{16}$ and $10^{724}$. 


\textbf{Options:} 
Each system features a distinct set of options that may influence performance, including categorical options (e.g., enumerated and binary) and numerical options (e.g., integers), as have been used in prior work~\cite{DBLP:journals/pvldb/ZhangCLWTLC22,zhang2021restune,DBLP:journals/pvldb/KanellisDKMCV22,he2022multi,chen2021efficient}.

\textbf{Fidelity:} 
To generate fidelity settings, we use: 



\begin{itemize}
    \item \textbf{\textsc{Sysbench}~\cite{sysbench}}: A transactional database benchmarking tool, generating stressed workloads controlled by various fidelity factors (e.g., \texttt{table-size} and \texttt{thread}).
    \item \textbf{\textsc{Wrk}~\cite{Wrk}}: A high-performance HTTP load generator for web servers, controlling request load and traffic intensity via fidelity factors such as \texttt{connections} and \texttt{duration}.
    \item \textbf{\textsc{Csmith}~\cite{Csmith}}: A $C$ program generator that produces programs of diverse scale and structural complexity for compilers, controlled by, e.g., \texttt{max-funcs} and \texttt{max-block-size}.
\end{itemize}

The fidelity used is summarized in Table~\ref{tb:workloads}. \revision{In practice, the perfect-fidelity setting is typically designated by developers based on business needs (often expensive), serving as the ground truth for assessing configuration quality.} To stress test the tuners, we set the setting that incurs the highest costs as the perfect-fidelity setting under which we seek to tune the configuration.

\begin{table}[t]
    \captionsetup{skip=8pt}
	\centering
	\tabcolsep=0.1 cm
	\caption{Subject systems studied. $\boldsymbol{\mathbfcal{X}}$ is the size of configuration space; ($|\boldsymbol{\mathbfcal{C}}|/|\boldsymbol{\mathbfcal{N}}|$) are the number of categorical/numerical options. tps/rps means transactions/requests per second.}
    \begin{adjustbox}{width=\linewidth,center}
	\label{tb:systems}
	\begin{threeparttable}
		
\begin{tabular}{llllllll}
\toprule
\textbf{System}  & \textbf{Domain} & \textbf{Performance} & \textbf{$|\boldsymbol{\mathbfcal{C}}|/|\boldsymbol{\mathbfcal{N}}|$} & \textbf{$\mathbfcal{X}$} &\textbf{Cost ($\boldsymbol{z}^*$)} & \textbf{Cost ($\boldsymbol{z}_{i}$)}\\ \midrule
\textsc{MySQL}~\cite{mysql_v}  & Database & Throughput (tps) & 2/18 & 4.24 $\times 10^{724}$ &$\approx$250s & 39s-90s\\
\rowcolor{teal!20} 
\textsc{PostgreSQL}~\cite{postgresql_v}  & Database & Throughput (tps) & 0/20 & 8.70 $\times 10^{122}$ &$\approx$250s &40s-76s\\
\textsc{Tomcat}~\cite{tomcat_v}  & Web server & Throughput (rps) & 3/17 & 4.47 $\times 10^{118}$ &$\approx$180s &10s-60s\\
\rowcolor{teal!20}
\textsc{Httpd}~\cite{httpd_v}  & Web server & Throughput (rps) & 7/13 & 4.76 $\times 10^{78}$ &$\approx$180s &10s-101s\\
\textsc{Gcc}~\cite{gcc_v}  & Compiler & Runtime (ms) & 20/0 & 7.21 $\times 10^{16}$ &$\approx$43s &2s-6s\\
\rowcolor{teal!20}
\textsc{Clang}~\cite{clang_v} & Compiler & Runtime (ms) & 20/0 & 7.21 $\times 10^{17}$ &$\approx$40s &2s-4s\\ \bottomrule
\end{tabular}

	\end{threeparttable}
    \end{adjustbox}
    
\end{table}




\begin{table}[t]
	\centering
    \captionsetup{skip=8pt}
	\caption{Details of fidelity generators. ``Type'' refers to the type of fidelity factors from \S\ref{subsection:fidelity_factors}. $\boldsymbol{z}^*$ and $\boldsymbol{\mathbfcal{Z}}$ denote the perfect-fidelity setting used and fidelity space, respectively.}
    \label{tb:workloads}
    \begin{adjustbox}{width=\linewidth,center}
	\begin{threeparttable}
		
\begin{tabular}{l|l|l|l|l|l}
\toprule
\multirow{2}{*}{\textbf{Benchmark}} & \multicolumn{2}{c|}{\textbf{Fidelity}} & \multirow{2}{*}{\textbf{Values}} & \multirow{2}{*}{\textbf{$\boldsymbol{\mathbfcal{Z}}$}} & \multirow{2}{*}{\textbf{$\boldsymbol{z}^{*}$}} \\ \cline{2-3}

& \textbf{Type} & \textbf{Factors} &  & &  \\ \midrule

\multirow{5}{*}{\textbf{\textsc{Sysbench}}} & Budget & \texttt{time} & $\{30, 35,..., 180\}$ & \multirow{5}{*}{$11200$} & $180$ \\
 & Dataset & \texttt{tables} & $\{20, 25,..., 50\}$ &  & $50$ \\
 & Budget & \texttt{threads} & $\{4, 6,..., 10\}$ &  & $4$ \\
 & Workload & \texttt{r/w ratio} & $\{0.5, 0.6,..., 0.9\}$ &  & $0.5$ \\
 & Dataset & \texttt{table-size} & $\{10K, 15K,..., 100K\}$ &  & $100K$ \\ \hline
\multirow{4}{*}{\textbf{\textsc{Wrk}}} & Workload & \texttt{post} & $\{true, false\}$ & \multirow{4}{*}{$1296$} & $true$ \\
 & Budget & \texttt{threads} & $\{1, 2,..., 10\}$ &  & $8$ \\
 & Budget & \texttt{duration} & $\{10, 15,..., 180\}$ &  & $180$ \\
 & Workload & \texttt{connections} & $\{10, 15,..., 50\}$ &  & $50$ \\ \hline
\multirow{5}{*}{\textbf{\textsc{Csmith}}} & Workload & \texttt{max-funcs} & $\{5, 10,..., 50\}$ & \multirow{5}{*}{$5880$} & $50$ \\
 & Dataset & \texttt{max-block-size} & $\{1, 2, 3, 4\}$ &  & $4$ \\
 & Workload & \texttt{max-block-depth} & $\{1, 2, 3, 6\}$ &  & $6$ \\
 & Workload & \texttt{inline-function-prob} & $\{10, 50,..., 90\}$ &  & $10$ \\
 & Workload & \texttt{max-array-len-per-dim} & $\{5, 10,..., 50\}$ &  & $50$ \\ \bottomrule
\end{tabular}

	\end{threeparttable}
    \end{adjustbox}
\end{table}

\subsection{State-of-the-Art Tuners}
\label{subsection:sota_tuners}
We compare \approach~against diverse types of tuners\footnote{\texttt{FLASH} uses exhaustive sampling at each iteration for acquisition evaluation, which does not work on systems with an intractable space; we extend this, namely \texttt{FLASH+}, by randomly sampling $1000$ configurations.}, spanning across different domains, as summarized in Table~\ref{tb:tuners}.

\begin{itemize}
    \item \textbf{Single-fidelity tuners} only conduct tuning under the perfect-fidelity setting.

    \item \revision{\textbf{Multi-fidelity tuners} exploit resource-allocation strategies across different fidelity settings. Since they work on a single fidelity factor and assume a monotonic relationship between cost and fidelity perfection, we preserve their original design by choosing the factor that is the most influential on cost.}
    
\end{itemize}

Regardless of the fidelity, a tuner can be either model-based or model-free: the former learns surrogate models (e.g., Gaussian Process) to approximate the configuration–performance landscape, steering tuning for the promising regions; the latter relies solely on system measurements to guide the tuning.

\subsection{Metrics}

The evaluation metrics would clearly be the performance metric that a system is of concerned, i.e., throughput or runtime. We also measure the relative efficiency and budget utilization of \approach~via $\Delta\mathcal{B}$: the budget saving (or overspend) for \approach~to reach the best performance of a counterpart tuner, if applicable, as $\Delta\mathcal{B} = \mathcal{B}_c - \mathcal{B}_{m}$, where $\mathcal{B}_c$ is the earliest time the counterpart tuner reaches its best performance (average over runs) while $\mathcal{B}_m$ is the time when \approach~achieves the same. As such, $\Delta\mathcal{B}>0$ indicates the amount of budget saved by \approach; or otherwise it means \approach~has no budget saving ($\Delta\mathcal{B}=0$) or could even overspend ($\Delta\mathcal{B}<0$). 

\subsection{Tuning Budget}

In this work, we set the tuning budget $\mathcal{B}$ as the wall-clock time allowed to tune the real system. To ensure realism of our experiment and consider the diverse measurement costs of different systems domains, we set $\mathcal{B}=24$ hours for \textsc{MySQL}/\textsc{PostgreSQL}, $\mathcal{B}=12$ hours for \textsc{Tomcat}/\textsc{Httpd}, and $\mathcal{B}=4$ hours for \textsc{Gcc}/\textsc{Clang}. These time budgets correspond to $\approx250$--$350$ measurements under the perfect-fidelity setting, which aligns with most of the prior studies~\cite{chen2025accuracy}. Note that we also examine the trajectories of tuning therein.

\begin{table}[t]
    \captionsetup{skip=8pt}
	\centering
	\small
	\tabcolsep=0.15 cm
	\caption{Summary of the compared tuners.}
    \begin{adjustbox}{width=\linewidth,center}
	\label{tb:tuners}
	\begin{threeparttable}

\begin{tabular}{lllll}
\toprule
\textbf{Tuner} & \textbf{Fidelity} & \textbf{Strategy} & \textbf{Domain} & \textbf{Year} \\ \midrule
\texttt{PromiseTune}~\cite{chen2025promisetune} & Single-fidelity & Model-based & Configuration & 2026 \\
\rowcolor{teal!20}
\texttt{HEBO}~\cite{cowen2022hebo} & Single-fidelity & Model-based & General & 2022 \\
\texttt{GA}~\cite{shahbazian2020equal} & Single-fidelity & Model-free & General & 2020 \\
\rowcolor{teal!20}
\texttt{FLASH+}~\cite{nair2018finding} & Single-fidelity & Model-based & Configuration & 2018 \\
\texttt{BestConfig}~\cite{zhu2017bestconfig} & Single-fidelity & Model-free & Configuration & 2017 \\
\rowcolor{teal!20}
\texttt{SMAC}~\cite{hutter2011sequential} & Single-fidelity & Model-free & General & 2011 \\ \midrule
\texttt{PriorBand}~\cite{mallik2023priorband} & Multi-fidelity & Model-free & HPO & 2023 \\
\rowcolor{teal!20}
\texttt{DEHB}~\cite{awad-ijcai21} & Multi-fidelity & Model-free & HPO & 2021 \\
\texttt{BOHB}~\cite{falkner2018bohb} & Multi-fidelity & Model-based & HPO & 2018 \\
\rowcolor{teal!20}
\texttt{Hyperband}~\cite{li2018hyperband} & Multi-fidelity & Model-free & HPO & 2018 \\ \bottomrule
\end{tabular}


	\end{threeparttable}
    \end{adjustbox}
\end{table}

\subsection{Parameter Settings}
\label{subsection:parameter_settings}
The initial sample size for model-based tuners (e.g., \texttt{SMAC}, \texttt{FLASH+}, and \texttt{HEBO}) is set to 30, following common practice in prior studies~\cite{nair2018finding,chen2024mmo}. For other parameters, such as the configuration population size ($n=20$) of \texttt{GA}, we adopt either the default setting or those reported in the literature to ensure fair comparison~\cite{li2018hyperband,falkner2018bohb,awad-ijcai21,mallik2023priorband,chen2025promisetune}. For \approach, the initial configuration sample size for discovering imperfect-fidelity settings $l$ and fidelity population size $m$ are both set to 10. {We use a smaller fidelity population ($m \textless n$) because quantifying a fidelity setting is more expensive, which requires measuring a batch of $l$ configurations to estimate its perfection level.} In particular, we set $\alpha= 0.5$---a generally best value (see \S\ref{subsection:sensitivity_analysis}). To avoid bias, we repeat 10 runs for each experiment.


\begin{table*}[t]
    \captionsetup{skip=8pt}
	\centering
	\footnotesize
	\tabcolsep=0.1 cm
	\caption{Comparing \approach~with state-of-the-art tuners over 10 runs. $\uparrow$ and $\downarrow$ denote maximized (throughput) and minimized (runtime) performance, respectively. $r$ denotes Scott-Knott ESD rank on performance. $\Delta\mathcal{B}\geq0$ and $\Delta\mathcal{B}<0$ refers to the budget saving and overspend of \approach~compared with a counterpart, respectively. \textcolor{red}{\ding{55}} indicates that \approach~fails to achieve the best of a counterpart when budget runs out. The best ranked tuner(s) for each system is highlighted in \colorbox{teal!20}{green}.}
    \label{tb:RQ2_Effectiveness}
    \begin{adjustbox}{width=\linewidth,center}
	\begin{threeparttable}

\begin{tabular}{l|clr|clr|clr|clr|clr|clr}
\toprule
\multicolumn{1}{c|}{} & \multicolumn{3}{c|}{\textbf{\textsc{MySQL} ($\uparrow$)}} & \multicolumn{3}{c|}{\textbf{\textsc{PostgreSQL} ($\uparrow$)}} & \multicolumn{3}{c|}{\textbf{\textsc{Httpd} ($\uparrow$)}} & \multicolumn{3}{c|}{\textbf{\textsc{Tomcat} ($\uparrow$)}} & \multicolumn{3}{c|}{\textbf{\textsc{Gcc} ($\downarrow$)}} & \multicolumn{3}{c}{\textbf{\textsc{Clang} ($\downarrow$)}} \\ \cline{2-19} 
\multicolumn{1}{c|}{\multirow{-2}{*}{\textbf{Tuner}}} & \textbf{$r$} & \textbf{Mean (Std)} & \textbf{$\Delta\mathcal{B}$} & \textbf{$r$} & \textbf{Mean (Std)} & \textbf{$\Delta\mathcal{B}$} & \textbf{$r$} & \textbf{Mean (Std)} & \textbf{$\Delta\mathcal{B}$} & \textbf{$r$} & \textbf{Mean (Std)} & \textbf{$\Delta\mathcal{B}$} & \textbf{$r$} & \textbf{Mean (Std)} & \textbf{$\Delta\mathcal{B}$} & \textbf{$r$} & \textbf{Mean (Std)} & \textbf{$\Delta\mathcal{B}$} \\ \midrule
\textbf{\texttt{Hyperband}} & 3 & 360.12 (20.61) & 9.7 hours & 7 & 498.71 (33.27) & 13.1 hours & 2 & 3028.27 (303.15) & 3.4 hours & 4 & 2890.92 (67.34) & 7.4 hours & 3 & 68.01 (1.98) & 0.6 hours & 5 & 69.09 (3.03) & 0.1 hours \\
\textbf{\texttt{BOHB}} & 3 & 358.35 (20.24) & 9.9 hours & 7 & 503.46 (34.03) & 23.6 hours & 6 & 2826.15 (62.30) & 6.7 hours & 5 & 2880.84 (87.21) & 6.6 hours & 5 & 69.85 (2.50) & 0.2 hours & \cellcolor{teal!20}1 & \cellcolor{teal!20}67.41 (2.83) & \cellcolor{teal!20}\textcolor{red}{\ding{55}} \\
\textbf{\texttt{DEHB}} & 2 & 365.50 (30.26) & 7.5 hours & 6 & 514.26 (68.39) & 21.7 hours & 5 & 2849.06 (80.17) & 6.7 hours & 7 & 2784.64 (50.90) & 6.9 hours & 4 & 69.65 (2.65) & 0.8 hours & 3 & 68.04 (3.40) & \textcolor{red}{\ding{55}} \\
\textbf{\texttt{PriorBand}} & \cellcolor{teal!20}1 & \cellcolor{teal!20}378.02 (22.10) & \cellcolor{teal!20}0.1 hours & 6 & 516.47 (37.93) & 17.9 hours & 2 & 3049.98 (578.41) & 3.0 hours & 4 & 2899.83 (60.98) & 6.8 hours & 4 & 69.62 (2.82) & 0.2 hours & 6 & 71.27 (2.76) & 0.4 hours \\
\textbf{\texttt{SMAC}} & 3 & 361.56 (31.48) & 9.5 hours & 2 & 556.77 (34.14) & 4.2 hours & 2 & 2969.58 (297.48) & 5.0 hours & 3 & 2926.93 (133.76) & 6.0 hours & 2 & 67.26 (1.84) & $-$0.9 hours & 2 & 67.63 (2.60) & \textcolor{red}{\ding{55}} \\
\textbf{\texttt{BestConfig}} & 2 & 365.06 (19.58) & 7.1 hours & 6 & 519.42 (35.35) & 5.2 hours & 6 & 2817.72 (55.43) & 7.2 hours & 6 & 2820.93 (69.37) & 6.4 hours & 4 & 69.29 (3.10) & 0.8 hours & 3 & 68.32 (2.23) & \textcolor{red}{\ding{55}} \\
\textbf{\texttt{FLASH+}} & \cellcolor{teal!20}1 & \cellcolor{teal!20}374.72 (14.79) & \cellcolor{teal!20}$-$0.5 hours & 3 & 541.14 (30.30) & 16.6 hours & 2 & 3010.90 (268.23) & 3.6 hours & \cellcolor{teal!20}1 & \cellcolor{teal!20}3309.26 (730.02) & \cellcolor{teal!20}1.9 hours & 2 & 67.92 (2.30) & 0.6 hours & 4 & 69.02 (1.90) & 0.0 hours \\
\textbf{\texttt{GA}} & 4 & 350.90 (30.00) & 6.3 hours & 3 & 542.72 (42.63) & 16.7 hours & 4 & 2877.19 (73.24) & 5.2 hours & 4 & 2898.54 (87.09) & 6.6 hours & 2 & 67.78 (2.91) & 0.5 hours & 4 & 68.69 (2.41) & $-$1.4 hours \\
\textbf{\texttt{HEBO}} & \cellcolor{teal!20}1 & \cellcolor{teal!20}378.03 (18.34) & \cellcolor{teal!20}0.0 hours & 4 & 537.17 (22.20) & 22.8 hours & 3 & 2920.09 (118.56) & 4.9 hours & 2 & 3072.72 (613.00) & 5.7 hours & 5 & 70.14 (2.34) & 0.7 hours & 4 & 69.07 (2.97) & $-$0.1 hours \\
\textbf{\texttt{PromiseTune}} & 2 & 365.24 (23.10) & 5.0 hours & 5 & 532.60 (14.34) & 22.8 hours & \cellcolor{teal!20}1 & \cellcolor{teal!20}3259.97 (532.48) & \cellcolor{teal!20}2.6 hours & 2 & 3145.03 (659.50) & 4.2 hours & 3 & 69.12 (1.84) & 0.9 hours & 6 & 70.26 (1.40) & 0.3 hours \\
\textbf{\approach} & \cellcolor{teal!20}1 & \cellcolor{teal!20}378.14 (21.75) & \cellcolor{teal!20}0.0 hours & \cellcolor{teal!20}1 & \cellcolor{teal!20}570.54 (37.20) & \cellcolor{teal!20}0.0 hours & \cellcolor{teal!20}1 & \cellcolor{teal!20}3359.75 (710.07) & \cellcolor{teal!20}0.0 hours & \cellcolor{teal!20}1 & \cellcolor{teal!20}3323.13 (843.83) & \cellcolor{teal!20}0.0 hours & \cellcolor{teal!20}1 & \cellcolor{teal!20}67.01 (2.93) & \cellcolor{teal!20}0.0 hours & 3 & 68.65 (4.43) & 0.0 hours \\ \bottomrule
\end{tabular}

	\end{threeparttable}
    \end{adjustbox}
\end{table*}

\subsection{Test for Statistical Significance}
To ensure statistical significance when comparing the performance of multiple tuners, we employ the Scott-Knott ESD test~\cite{tantithamthavorn2016empirical}. In a nutshell, it begins by ranking tuners according to their mean performance, then recursively partitions this ordered list into statistically different subgroups. For instance, given three tuners $A$, $B$, and $C$, Scott-Knott ESD might partition them into two subgroups: $\{A, B\}$ at rank $r=1$ and $\{C\}$ at rank $r=2$. This indicates that $A$ and $B$ cannot be distinguished statistically, yet both significantly outperform $C$. Scott-Knott ESD is chosen because, unlike the Kruskal-Wallis test, it overcomes the confounding factor of overlapping groups and eliminates the dependence on post-hoc correlation~\cite{mchugh2011multiple}.



\section{Results and Analysis}
\label{section:results_and_analysis}

\subsection{RQ1: Benefits of Imperfect-Fidelity}





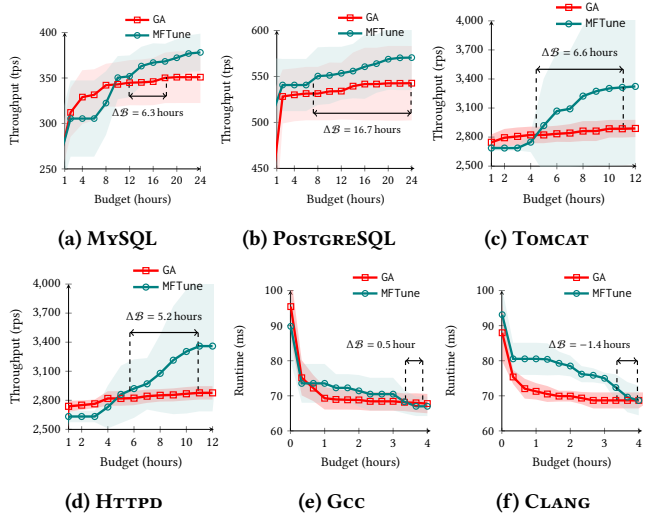
\begin{figure}
    \subfloat[\textsc{MySQL}\label{fig:rq1_mysql}]{\begin{tikzpicture}
    \begin{axis}[
      axis x line  = bottom,
            axis y line  = left  ,
        width=5cm, height=5cm, 
        xlabel={Budget (hours)},
        ylabel={Throughput (tps)},
        xmin=1, xmax=24.1,
        xtick={1,4,8,12,16,20,24},
        ymin=250, ymax=400,
        ytick={250,300,350,400,450,500},
            legend cell align=left,
          legend columns=1,
        legend style={
          draw=none, fill=none,
          at={(0.98,0.9)}, anchor=south east,
          legend cell align=left
        }
    ]

\addplot[
    red,
    mark=square,mark options={thick,solid},
    ultra thick, forget plot
] coordinates {
(0.07, 256.75) 
(2.00, 312.12) 
(4.00, 328.99) 
(6.00, 331.53) 
(8.00, 342.09) 
(10.00, 342.99) 
(12.00, 344.87) 
(14.00, 345.16) 
(16.00, 346.05) 
(18.00, 350.18) 
(20.00, 350.90) 
(22.00, 350.90) 
(24.00, 350.90) 
    }; 
\addlegendentry{\texttt{GA}}

\addplot[
name path=upper_low,
draw=none, forget plot
] coordinates {
(0.07, 277.48) 
(2.00, 339.01) 
(4.00, 358.58) 
(6.00, 359.59) 
(8.00, 365.67) 
(10.00, 367.55) 
(12.00, 369.67) 
(14.00, 370.41) 
(16.00, 372.03) 
(18.00, 377.96) 
(20.00, 379.36) 
(22.00, 379.36) 
(24.00, 379.36) 
}; 

\addplot[
name path=lower_low,
draw=none, forget plot
] coordinates {
(0.07, 236.02) (2.00, 285.23) (4.00, 299.39) (6.00, 303.46) (8.00, 318.50) (10.00, 318.43) (12.00, 320.07) (14.00, 319.92) (16.00, 320.07) (18.00, 322.41) (20.00, 322.44) (22.00, 322.44) (24.00, 322.44)
}; 
\addplot[
    red!30,
    fill opacity=0.30, forget plot
] fill between[
    of=upper_low and lower_low,
];

\addplot[
teal,
mark=o,mark options={thick},
ultra thick, forget plot
] coordinates {
(0.07, 259.69) (2.00, 305.37) (4.00, 305.37) (6.00, 305.37) (8.00, 322.41) (10.00, 350.39) (12.00, 351.87) (14.00, 363.10) (16.00, 366.97) (18.00, 368.25) (20.00, 372.25) (22.00, 376.92) (24.00, 378.14)
}; 
\addlegendentry{\texttt{MFTune}}

\addplot[
name path=upper_low,
draw=none, forget plot
] coordinates {
(0.07, 298.94) (2.00, 347.30) (4.00, 347.30) (6.00, 347.30) (8.00, 357.47) (10.00, 374.13) (12.00, 374.86) (14.00, 384.37) (16.00, 388.18) (18.00, 387.64) (20.00, 392.70) (22.00, 397.00) (24.00, 398.78) 
}; 

\addplot[
name path=lower_low,
draw=none, forget plot
] coordinates {
(0.07, 220.44) (2.00, 263.43) (4.00, 263.43) (6.00, 263.43) (8.00, 287.35) (10.00, 326.65) (12.00, 328.89) (14.00, 341.83) (16.00, 345.77) (18.00, 348.86) (20.00, 351.80) (22.00, 356.84) (24.00, 357.51) 
}; 

\addplot[
    teal!30,
    fill opacity=0.30, forget plot
] fill between[
    of=upper_low and lower_low,
];

\addlegendimage{red, mark=square, mark options={thick,solid}, ultra thick}
\addlegendentry{\texttt{GA}}

\addlegendimage{teal, mark=o, mark options={thick}, ultra thick}
\addlegendentry{\texttt{MFTune}}


\def\Bmf{11.922}

\def\Bga{18.238}

\def\ArrowY{330}

\draw[dashed, thick]
(axis cs:\Bmf, \ArrowY) -- (axis cs:\Bmf, 350.90);

\draw[dashed, thick]
(axis cs:\Bga, \ArrowY) -- (axis cs:\Bga, 350.90);

\draw[<->, thick]
(axis cs:\Bmf, \ArrowY)
--
(axis cs:\Bga, \ArrowY)
node[midway, below, yshift=-5pt]
{\small $\Delta\mathcal{B} = 6.3\,\mathrm{hours}$};

\end{axis}

\end{tikzpicture}}
    \hfill  
    \subfloat[\textsc{PostgreSQL}\label{fig:rq1_mysql}]{\begin{tikzpicture}
    \begin{axis}[
      axis x line  = bottom,
            axis y line  = left  ,
        width=5cm, height=5cm, 
        xlabel={Budget (hours)},
        ylabel={Throughput (tps)},
        xmin=1, xmax=24.5,
        xtick={1,4,8,12,16,20,24},
        ymin=450, ymax=600,
        ytick={450,500,550,600,650},
            legend cell align=left,
          legend columns=1,
        legend style={
          draw=none, fill=none,
          at={(0.98,0.9)}, anchor=south east,
          legend cell align=left
        }
    ]

\addplot[
    red,
    mark=square,mark options={thick,solid},
    ultra thick, forget plot
] coordinates {
(0.07, 423.33) (2.00, 528.25) (4.00, 530.08) (6.00, 531.40) (8.00, 531.40) (10.00, 533.78) (12.00, 534.09) (14.00, 539.48) (16.00, 541.76) (18.00, 541.76) (20.00, 542.56) (22.00, 542.56) (24.00, 542.72) 
    }; 
\addlegendentry{\texttt{GA}}

\addplot[
name path=upper_low,
draw=none, forget plot
] coordinates {
(0.07, 526.27) (2.00, 557.58) (4.00, 558.86) (6.00, 560.44) (8.00, 560.44) (10.00, 565.20) (12.00, 564.99) (14.00, 576.11) (16.00, 581.24) (18.00, 581.24) (20.00, 583.22) (22.00, 583.22) (24.00, 583.16)  
}; 

\addplot[
name path=lower_low,
draw=none, forget plot
] coordinates {
(0.07, 320.39) (2.00, 498.91) (4.00, 501.29) (6.00, 502.37) (8.00, 502.37) (10.00, 502.35) (12.00, 503.19) (14.00, 502.85) (16.00, 502.28) (18.00, 502.28) (20.00, 501.91) (22.00, 501.91) (24.00, 502.27) 
}; 
\addplot[
    red!30,
    fill opacity=0.30, forget plot
] fill between[
    of=upper_low and lower_low,
];

\addplot[
teal,
mark=o,mark options={thick},
ultra thick, forget plot
] coordinates {
(0.07, 506.21) (2.00, 540.72) (4.00, 540.72) (6.00, 540.72) (8.00, 550.42) (10.00, 551.21) (12.00, 553.52) (14.00, 555.89) (16.00, 560.64) (18.00, 563.99) (20.00, 568.80) (22.00, 570.34) (24.00, 570.54) 
}; 
\addlegendentry{\texttt{MFTune}}

\addplot[
name path=upper_low,
draw=none, forget plot
] coordinates {
(0.07, 570.05) (2.00, 569.09) (4.00, 569.09) (6.00, 569.09) (8.00, 581.79) (10.00, 582.25) (12.00, 582.28) (14.00, 585.48) (16.00, 595.57) (18.00, 595.70) (20.00, 602.04) (22.00, 605.52) (24.00, 605.83) 
}; 

\addplot[
name path=lower_low,
draw=none, forget plot
] coordinates {
(0.07, 442.37) (2.00, 512.34) (4.00, 512.34) (6.00, 512.34) (8.00, 519.05) (10.00, 520.17) (12.00, 524.75) (14.00, 526.29) (16.00, 525.71) (18.00, 532.28) (20.00, 535.56) (22.00, 535.16) (24.00, 535.25) 
}; 

\addplot[
    teal!30,
    fill opacity=0.30, forget plot
] fill between[
    of=upper_low and lower_low,
];

\addlegendimage{red, mark=square, mark options={thick,solid}, ultra thick}
\addlegendentry{\texttt{GA}}

\addlegendimage{teal, mark=o, mark options={thick}, ultra thick}
\addlegendentry{\texttt{MFTune}}

\def\Bmf{7.237}

\def\Bga{23.887}

\def\ArrowY{510}

\draw[dashed, thick]
(axis cs:\Bmf, \ArrowY) -- (axis cs:\Bmf, 542.72);

\draw[dashed, thick]
(axis cs:\Bga, \ArrowY) -- (axis cs:\Bga, 542.72);

\draw[<->, thick]
(axis cs:\Bmf, \ArrowY)
--
(axis cs:\Bga, \ArrowY)
node[midway, below, yshift=-5pt]
{\small $\Delta\mathcal{B} = 16.7\,\mathrm{hours}$};

\end{axis}

\end{tikzpicture}}
    \hfill  
    \subfloat[\textsc{Tomcat}\label{fig:rq1_tomcat}]{\begin{tikzpicture}
    \begin{axis}[
      axis x line  = bottom,
            axis y line  = left  ,
        width=5cm, height=5cm, 
        xlabel={Budget (hours)},
        ylabel={Throughput (rps)},
        xmin=1, xmax=12.1,
        xtick={1,2,4,6,8,10,12},
        ymin=2500, ymax=4000,
        ytick={2500,2800,3100,3400,3700,4000},
            legend cell align=left,
          legend columns=1,
        legend style={
          draw=none, fill=none,
          at={(0.98,0.9)}, anchor=south east,
          legend cell align=left
        }
    ]

\addplot[
    red,
    mark=square,mark options={thick,solid},
    ultra thick, forget plot
] coordinates {
(0.05, 2623.82) (1.00, 2746.36) (2.00, 2793.60) (3.00, 2807.07) (4.00, 2822.35) (5.00, 2822.37) (6.00, 2834.33) (7.00, 2841.68) (8.00, 2862.22) (9.00, 2862.45) (10.00, 2886.21) (11.00, 2886.21) (12.00, 2889.11)

    }; 
\addlegendentry{\texttt{GA}}

\addplot[
name path=upper_low,
draw=none, forget plot
] coordinates {
(0.05, 2657.30) (1.00, 2827.63) (2.00, 2884.29) (3.00, 2899.48) (4.00, 2907.90) (5.00, 2907.89) (6.00, 2917.61) (7.00, 2933.95) (8.00, 2965.36) (9.00, 2965.79) (10.00, 2977.36) (11.00, 2977.36) (12.00, 2978.32) 
}; 

\addplot[
name path=lower_low,
draw=none, forget plot
] coordinates {
(0.05, 2590.33) (1.00, 2665.09) (2.00, 2702.92) (3.00, 2714.67) (4.00, 2736.79) (5.00, 2736.86) (6.00, 2751.04) (7.00, 2749.41) (8.00, 2759.08) (9.00, 2759.11) (10.00, 2795.07) (11.00, 2795.07) (12.00, 2799.90) 
}; 
\addplot[
    red!70,
    fill opacity=0.30, forget plot
] fill between[
    of=upper_low and lower_low,
];

\addplot[
teal,
mark=o,mark options={thick},
ultra thick, forget plot
] coordinates {
(0.05, 2640.71) (1.00, 2686.96) (2.00, 2686.96) (3.00, 2686.96) (4.00, 2747.67) (5.00, 2918.63) (6.00, 3068.71) (7.00, 3091.45) (8.00, 3223.29) (9.00, 3273.12) (10.00, 3302.19) (11.00, 3313.21) (12.00, 3323.03)

}; 
\addlegendentry{\texttt{MFTune}}

\addplot[
name path=upper_low,
draw=none, forget plot
] coordinates {
(0.05, 2688.87) (1.00, 2712.76) (2.00, 2712.76) (3.00, 2712.76) (4.00, 2850.21) (5.00, 3403.72) (6.00, 3680.87) (7.00, 3691.44) (8.00, 3950.06) (9.00, 4039.94) (10.00, 4102.05) (11.00, 4118.10) (12.00, 4123.62) 
}; 

\addplot[
name path=lower_low,
draw=none, forget plot
] coordinates {
(0.05, 2592.55) (1.00, 2661.17) (2.00, 2661.17) (3.00, 2661.17) (4.00, 2645.13) (5.00, 2433.54) (6.00, 2456.56) (7.00, 2491.46) (8.00, 2496.53) (9.00, 2506.30) (10.00, 2502.34) (11.00, 2508.32) (12.00, 2522.45) 
}; 

\addplot[
    teal!30,
    fill opacity=0.30, forget plot
] fill between[
    of=upper_low and lower_low,
];

\addlegendimage{red, mark=square, mark options={thick,solid}, ultra thick}
\addlegendentry{\texttt{GA}}

\addlegendimage{teal, mark=o, mark options={thick}, ultra thick}
\addlegendentry{\texttt{MFTune}}

\def\Bmf{4.426}

\def\Bga{11.074}

\def\ArrowY{3500}

\draw[dashed, thick]
(axis cs:\Bmf, \ArrowY) -- (axis cs:\Bmf, 2889.11);

\draw[dashed, thick]
(axis cs:\Bga, \ArrowY) -- (axis cs:\Bga, 2889.11);

\draw[<->, thick]
(axis cs:\Bmf, \ArrowY)
--
(axis cs:\Bga, \ArrowY)
node[midway, above, yshift=5pt]
{\small $\Delta\mathcal{B} = 6.6\,\mathrm{hours}$};

\end{axis}

\end{tikzpicture}}

    \vspace{0.05cm}

    \subfloat[\textsc{Httpd}\label{fig:rq1_httpd}]{\begin{tikzpicture}
    \begin{axis}[
      axis x line  = bottom,
            axis y line  = left  ,
        width=5cm, height=5cm, 
        xlabel={Budget (hours)},
        ylabel={Throughput (rps)},
        xmin=1, xmax=12.1,
        xtick={1,2,4,6,8,10,12},
        ymin=2500, ymax=4000,
        ytick={2500,2800,3100,3400,3700,4000},
            legend cell align=left,
          legend columns=1,
        legend style={
          draw=none, fill=none,
          at={(0.98,0.9)}, anchor=south east,
          legend cell align=left
        }
    ]

\addplot[
    red,
    mark=square,mark options={thick,solid},
    ultra thick, forget plot
] coordinates {
(0.05, 1104.69) (1.00, 2738.10) (2.00, 2750.41) (3.00, 2763.90) (4.00, 2819.36) (5.00, 2819.36) (6.00, 2822.64) (7.00, 2842.72) (8.00, 2851.73) (9.00, 2856.07) (10.00, 2867.72) (11.00, 2877.19) (12.00, 2877.19) 
    }; 
\addlegendentry{\texttt{GA}}

\addplot[
name path=upper_low,
draw=none, forget plot
] coordinates {
(0.05, 1952.64) (1.00, 2794.51) (2.00, 2799.03) (3.00, 2808.21) (4.00, 2894.75) (5.00, 2894.75) (6.00, 2895.87) (7.00, 2911.91) (8.00, 2916.50) (9.00, 2923.65) (10.00, 2937.22) (11.00, 2946.67) (12.00, 2946.67) 
}; 

\addplot[
name path=lower_low,
draw=none, forget plot
] coordinates {
(0.05, 256.75) (1.00, 2681.68) (2.00, 2701.80) (3.00, 2719.59) (4.00, 2743.97) (5.00, 2743.97) (6.00, 2749.41) (7.00, 2773.53) (8.00, 2786.97) (9.00, 2788.48) (10.00, 2798.22) (11.00, 2807.71) (12.00, 2807.71) 
}; 
\addplot[
    red!70,
    fill opacity=0.30, forget plot
] fill between[
    of=upper_low and lower_low,
];

\addplot[
teal,
mark=o,mark options={thick},
ultra thick, forget plot
] coordinates {
(0.06, 1888.43) (1.00, 2633.05) (2.00, 2633.05) (3.00, 2633.05) (4.00, 2730.65) (5.00, 2861.11) (6.00, 2921.08) (7.00, 2970.38) (8.00, 3078.34) (9.00, 3214.38) (10.00, 3303.88) (11.00, 3359.75) (12.00, 3359.75) 
}; 
\addlegendentry{\texttt{MFTune}}

\addplot[
name path=upper_low,
draw=none, forget plot
] coordinates {
(0.06, 2788.27) (1.00, 2692.18) (2.00, 2692.18) (3.00, 2692.18) (4.00, 2951.15) (5.00, 3157.36) (6.00, 3223.14) (7.00, 3236.08) (8.00, 3477.91) (9.00, 3771.70) (10.00, 3943.49) (11.00, 4033.38) (12.00, 4033.38) (13.00, 4033.38) 
}; 

\addplot[
name path=lower_low,
draw=none, forget plot
] coordinates {
(0.06, 988.59) (1.00, 2573.92) (2.00, 2573.92) (3.00, 2573.92) (4.00, 2510.14) (5.00, 2564.87) (6.00, 2619.02) (7.00, 2704.67) (8.00, 2678.77) (9.00, 2657.06) (10.00, 2664.26) (11.00, 2686.12) (12.00, 2686.12)
}; 

\addplot[
    teal!30,
    fill opacity=0.30, forget plot
] fill between[
    of=upper_low and lower_low,
];

\addlegendimage{red, mark=square, mark options={thick,solid}, ultra thick}
\addlegendentry{\texttt{GA}}

\addlegendimage{teal, mark=o, mark options={thick}, ultra thick}
\addlegendentry{\texttt{MFTune}}

\def\Bmf{5.686}

\def\Bga{10.9}

\def\ArrowY{3500}

\draw[dashed, thick]
(axis cs:\Bmf, \ArrowY) -- (axis cs:\Bmf, 2877.19);

\draw[dashed, thick]
(axis cs:\Bga, \ArrowY) -- (axis cs:\Bga, 2877.19);

\draw[<->, thick]
(axis cs:\Bmf, \ArrowY)
--
(axis cs:\Bga, \ArrowY)
node[midway, above, yshift=5pt]
{\small $\Delta\mathcal{B} = 5.2\,\mathrm{hours}$};

\end{axis}

\end{tikzpicture}}
    \hfill  
        \subfloat[\textsc{Gcc}\label{fig:rq1_gcc}]{\begin{tikzpicture}
    \begin{axis}[
      axis x line  = bottom,
            axis y line  = left  ,
        width=5cm, height=5cm, 
        xlabel={Budget (hours)},
        ylabel={Runtime (ms)},
        xmin=0, xmax=4.1,
        xtick={0,1,2,3,4},
        ymin=60, ymax=100,
        ytick={60,70,80,90,100},
            legend cell align=left,
          legend columns=1,
        legend style={
          draw=none, fill=none,
          at={(0.98,0.9)}, anchor=south east,
          legend cell align=left
        }
    ]

\addplot[
    red,
    mark=square,mark options={thick,solid},
    ultra thick, forget plot
] coordinates {
(0.01, 95.42) (0.33, 75.13) (0.67, 72.20) (1.00, 69.31) (1.33, 68.96) (1.67, 68.86) (2.00, 68.82) (2.33, 68.42) (2.67, 68.42) (3.00, 68.42) (3.33, 68.16) (3.67, 67.99) (4.00, 67.78)

    }; 
\addlegendentry{\texttt{GA}}

\addplot[
name path=upper_low,
draw=none, forget plot
] coordinates {
(0.01, 104.87) (0.33, 80.16) (0.67, 75.91) (1.00, 72.53) (1.33, 71.85) (1.67, 71.66) (2.00, 71.65) (2.33, 71.10) (2.67, 71.10) (3.00, 71.10) (3.33, 70.67) (3.67, 70.71) (4.00, 70.54) 
}; 

\addplot[
name path=lower_low,
draw=none, forget plot
] coordinates {
(0.01, 85.97) (0.33, 70.11) (0.67, 68.48) (1.00, 66.09) (1.33, 66.07) (1.67, 66.07) (2.00, 65.99) (2.33, 65.75) (2.67, 65.75) (3.00, 65.75) (3.33, 65.64) (3.67, 65.27) (4.00, 65.02) 
}; 
\addplot[
    red!70,
    fill opacity=0.30, forget plot
] fill between[
    of=upper_low and lower_low,
];

\addplot[
teal,
mark=o,mark options={thick},
ultra thick, forget plot
] coordinates {
(0.01, 89.79) (0.33, 73.54) (0.67, 73.54) (1.00, 73.54) (1.33, 72.27) (1.67, 72.27) (2.00, 71.41) (2.33, 70.48) (2.67, 70.48) (3.00, 70.48) (3.33, 68.29) (3.67, 67.01) (4.00, 67.01)

}; 
\addlegendentry{\texttt{MFTune}}

\addplot[
name path=upper_low,
draw=none, forget plot
] coordinates {
(0.01, 99.93) (0.33, 79.07) (0.67, 79.07) (1.00, 79.07) (1.33, 76.18) (1.67, 76.18) (2.00, 75.92) (2.33, 75.13) (2.67, 75.13) (3.00, 75.13) (3.33, 71.59) (3.67, 69.79) (4.00, 69.79) 
}; 

\addplot[
name path=lower_low,
draw=none, forget plot
] coordinates {
(0.01, 79.66) (0.33, 68.01) (0.67, 68.01) (1.00, 68.01) (1.33, 68.36) (1.67, 68.36) (2.00, 66.90) (2.33, 65.82) (2.67, 65.82) (3.00, 65.82) (3.33, 64.99) (3.67, 64.23) (4.00, 64.23) 
}; 

\addplot[
    teal!30,
    fill opacity=0.30, forget plot
] fill between[
    of=upper_low and lower_low,
];

\addlegendimage{red, mark=square, mark options={thick,solid}, ultra thick}
\addlegendentry{\texttt{GA}}

\addlegendimage{teal, mark=o, mark options={thick}, ultra thick}
\addlegendentry{\texttt{MFTune}}

\def\Bmf{3.362}

\def\Bga{3.865}

\def\ArrowY{80}

\draw[dashed, thick]
(axis cs:\Bmf, \ArrowY) -- (axis cs:\Bmf, 67.78);

\draw[dashed, thick]
(axis cs:\Bga, \ArrowY) -- (axis cs:\Bga, 67.78);

\draw[<->, thick]
(axis cs:\Bmf, \ArrowY)
--
(axis cs:\Bga, \ArrowY)
node[midway, above, yshift=5pt, xshift=-23pt]
{\small $\Delta\mathcal{B} = 0.5\,\mathrm{hour}$};

\end{axis}

\end{tikzpicture}}
    \hfill  
        \subfloat[\textsc{Clang}\label{fig:rq1_clang}]{\begin{tikzpicture}
    \begin{axis}[
      axis x line  = bottom,
            axis y line  = left  ,
        width=5cm, height=5cm, 
        xlabel={Budget (hours)},
        ylabel={Runtime (ms)},
        xmin=0, xmax=4.1,
        xtick={0,1,2,3,4},
        ymin=60, ymax=100,
        ytick={60,70,80,90,100},
        legend cell align=left,
        legend columns=1,
        legend style={
          draw=none, fill=none,
          at={(0.98,0.9)}, anchor=south east,
          legend cell align=left
        }
    ]

\addplot[
    red,
    mark=square,mark options={thick,solid},
    ultra thick, forget plot
] coordinates {
(0.01, 87.97) (0.33, 75.37) (0.67, 72.03) (1.00, 71.27) (1.33, 70.51) (1.67, 69.91) (2.00, 69.91) (2.33, 69.35) (2.67, 68.69) (3.00, 68.69) (3.33, 68.69) (3.67, 68.69) (4.00, 68.69)

    }; 
\addlegendentry{\texttt{GA}}

\addplot[
name path=upper_low,
draw=none, forget plot
] coordinates {
(0.01, 96.08) (0.33, 77.39) (0.67, 74.86) (1.00, 73.62) (1.33, 72.78) (1.67, 71.44) (2.00, 71.44) (2.33, 71.10) (2.67, 70.97) (3.00, 70.97) (3.33, 70.97) (3.67, 70.97) (4.00, 70.97) (4.33, 70.97) (4.67, 70.97)
}; 

\addplot[
name path=lower_low,
draw=none, forget plot
] coordinates {
(0.01, 79.86) (0.33, 73.35) (0.67, 69.20) (1.00, 68.92) (1.33, 68.24) (1.67, 68.37) (2.00, 68.37) (2.33, 67.60) (2.67, 66.40) (3.00, 66.40) (3.33, 66.40) (3.67, 66.40) (4.00, 66.40) 
}; 
\addplot[
    red!70,
    fill opacity=0.30, forget plot
] fill between[
    of=upper_low and lower_low,
];

\addplot[
teal,
mark=o,mark options={thick},
ultra thick, forget plot
] coordinates {
(0.01, 93.12) (0.33, 80.53) (0.67, 80.53) (1.00, 80.53) (1.33, 80.39) (1.67, 79.26) (2.00, 78.49) (2.33, 76.17) (2.67, 75.84) (3.00, 74.99) (3.33, 72.36) (3.67, 69.62) (4.00, 68.65)

}; 
\addlegendentry{\texttt{MFTune}}

\addplot[
name path=upper_low,
draw=none, forget plot
] coordinates {
(0.01, 100.58) (0.33, 85.11) (0.67, 85.11) (1.00, 85.11) (1.33, 84.96) (1.67, 82.60) (2.00, 81.44) (2.33, 78.47) (2.67, 77.85) (3.00, 76.61) (3.33, 76.10) (3.67, 73.91) (4.00, 72.85) 
}; 

\addplot[
name path=lower_low,
draw=none, forget plot
] coordinates {
(0.01, 85.66) (0.33, 75.94) (0.67, 75.94) (1.00, 75.94) (1.33, 75.82) (1.67, 75.92) (2.00, 75.54) (2.33, 73.87) (2.67, 73.84) (3.00, 73.37) (3.33, 68.61) (3.67, 65.32) (4.00, 64.44) 
}; 

\addplot[
    teal!30,
    fill opacity=0.30, forget plot
] fill between[
    of=upper_low and lower_low,
];

\addlegendimage{red, mark=square, mark options={thick,solid}, ultra thick}
\addlegendentry{\texttt{GA}}

\addlegendimage{teal, mark=o, mark options={thick}, ultra thick}
\addlegendentry{\texttt{MFTune}}

\def\Bmf{3.362}

\def\Bga{3.961}

\def\ArrowY{80}

\draw[dashed, thick]
(axis cs:\Bmf, \ArrowY) -- (axis cs:\Bmf, 68.69);

\draw[dashed, thick]
(axis cs:\Bga, \ArrowY) -- (axis cs:\Bga, 68.69);

\draw[<->, thick]
(axis cs:\Bmf, \ArrowY)
--
(axis cs:\Bga, \ArrowY)
node[midway, above, yshift=5pt, xshift=-26pt]
{\small $\Delta\mathcal{B} = -1.4\,\mathrm{hours}$};

\end{axis}

\end{tikzpicture}}
        
    \caption{Tuning trajectories with and without multi-fidelity over 10 runs.}
    \label{fig:RQ1_Curve}
    \vspace{-10pt}
\end{figure}

\subsubsection{Method.} For \textbf{RQ1}, we compare \approach~with a classic \texttt{GA}, which is essentially the single fidelity version of \approach~that tunes under the perfect-fidelity setting only; all other designs are identical. We plot the trajectory of the best configuration found under the perfect-fidelity setting (the ones in archive for \approach).

\subsubsection{Results.}
As shown in Figure~\ref{fig:RQ1_Curve}, \approach~consistently outperforms \texttt{GA} across all six systems, demonstrating superior performance and competitive convergence speed. In efficacy, \approach\ achieves up to 16.77\% performance improvement (on \textsc{Httpd}), suggesting that integrating the imperfect-fidelity setting enables more effective exploration of high-quality configurations for the perfect-fidelity setting. 

Notably, at the early stage, \approach~shows slightly slower convergence since part of its budget is devoted to discover the fair imperfect-fidelity setting. However, this initial investment quickly pays off as the subsequent tuning proceeds more efficiently. Once an appropriate fair imperfect-fidelity setting is found, the tuners benefits from it to accelerate convergence toward promising configurations under the perfect-fidelity setting. As a result, \approach~generally leads to significant budget utilization, saving up to $\Delta\mathcal{B}=24-7.3=16.7$ hours (on \textsc{PostgreSQL}).
Thus, we conclude that:
\begin{quotebox}
   \noindent
   \textit{\textbf{RQ1:} Considering imperfect fidelity settings in \approach~is greatly beneficial, yielding up to $16.77\%$ performance improvement and $16.7$ hours budget saving.}
\end{quotebox}

\subsection{RQ2: Effectiveness and Efficiency}

\subsubsection{Method.} For \textbf{RQ2}, we compare \approach~with 10 state-of-the-art tuners on all systems. To ensure statistical significance, we use the Scott-Knott ESD test to rank the tuners over 10 runs.

\subsubsection{Results.} As from Table~\ref{tb:RQ2_Effectiveness}, \approach~achieves remarkable results: it is ranked the best in 83.33\% (5/6) of the cases. This considerably outperforms the generally second-best tuner, \texttt{FLASH+}, which is ranked the best for 33.33\% (2/6) cases only. Among single-fidelity tuners, model-based approaches (e.g., \texttt{SMAC, FLASH+, HEBO,} and \texttt{PromiseTune}) generally outperform model-free ones (e.g., \texttt{GA} and \texttt{BestConfig}), as their surrogate models better balance exploration and exploitation, steering the tuning towards higher-quality configurations. Notably, multi-fidelity HPO tuners (e.g., \texttt{DEHB}) perform poorly. This is because they assume a monotonic relationship between evaluation cost and fidelity perfection, which is commonly not held for configurable systems (as in \S\ref{subsection:motivation_challenges}). In contrast, \approach’s handling of the interaction between multiple fidelity factors, together with the tuning for ``wideness'' and ``depth'',  has resulted in up to 19.34\% performance improvement (on \textsc{Tomcat}). 


\approach\ also shows clear practical advantages in budget utilization: it saves more budget than its counterparts in 50 out of 60 comparisons, with up to 23.6 hours. Therefore:


\begin{quotebox}
   \noindent
   \textit{\textbf{RQ2:} \approach~leads to considerably better tuning quality than others: it ranks first in 83.33\% of the cases with up to 19.34\% performance improvement, achieves better budget utilization in 50 comparisons, and saves up to 23.6 hours.}
\end{quotebox}

\subsection{RQ3: Ablation Study}

\subsubsection{Method.} For \textbf{RQ3}, we ablate \approach~with two variants:
\begin{itemize}
    \item \textbf{\approach-\texttt{I}} removes the first phase of imperfect-fidelity discovering and randomly select the fair imperfect-fidelity.
    \item \textbf{\approach-\texttt{II}} simplifies the two-stage tuning under the fair imperfect-fidelity setting, leaving only one stage without the progressive diversity sampling. 
\end{itemize}

Again, we apply the Scott-Knott test on all comparisons. 

\subsubsection{Results.} As summarized in Table~\ref{tb:RQ4_Ablation}, \approach~obtains the best rank in the majority of the cases (5/6), suggesting the necessity of both. Indeed, omitting imperfect-fidelity discovery might lead to a rather harmful imperfect-fidelity setting being used; simplifying two-stage tuning under imperfect-fidelity might cause the tuning to be prone to local-optimal, since without the global diversity encouragement, the ``wideness'' of tuning would be restricted. In light of the above, we can conclude:


\begin{quotebox}
   \noindent
   \textit{\textbf{RQ3:} Discovering imperfect-fidelity and its two-stage tuning are both essential to the superiority of \approach.}
\end{quotebox}

\begin{table}[t]
	\centering
	\footnotesize
	\tabcolsep=0.05 cm
	\caption{Comparing \approach~with its two variants over 10 runs. The format is the same as Table~\ref{tb:RQ2_Effectiveness}.} 
    \label{tb:RQ4_Ablation}
    \begin{adjustbox}{width=\linewidth,center}
	\begin{threeparttable}
		
\begin{tabular}{l|llr|llr||ll}
\toprule
\multicolumn{1}{c|}{} & \multicolumn{3}{c|}{\textbf{\approach-\texttt{I}}} & \multicolumn{3}{c||}{\textbf{\approach-\texttt{II}}} & \multicolumn{2}{c}{\textbf{\approach}} \\ \cline{2-9} 
\multicolumn{1}{l|}{\multirow{-2}{*}{\textbf{System}}} & \textbf{$r$} & \textbf{Mean (Std)} & \textbf{$\Delta\mathcal{B}$} & \textbf{$r$} & \textbf{Mean (Std)} & \textbf{$\Delta\mathcal{B}$} & \textbf{$r$} & \textbf{Mean (Std)}  \\ \midrule
\textbf{\textsc{MySQL}} & \cellcolor{teal!20}1 & \cellcolor{teal!20}375.53 (17.12) & \cellcolor{teal!20}1.7 hours & 2 & 338.17 (34.69) & 13.1 hours & \cellcolor{teal!20}1 & \cellcolor{teal!20}378.14 (21.75) \\
\textbf{\textsc{PostgreSQL}} & \cellcolor{teal!20}1 & \cellcolor{teal!20}601.17 (30.57) & \cellcolor{teal!20}\textcolor{red}{\ding{55}} & 3 & 552.67 (44.61) & 13.3 hours & 2 & 570.54 (37.20) \\
\textbf{\textsc{Tomcat}} & 2 & 2977.38 (271.63) & 7.3 hours & 3 & 2810.84 (63.48) & 6.6 hours & \cellcolor{teal!20}1 & \cellcolor{teal!20}3323.13 (843.83) \\
\textbf{\textsc{Httpd}} & 2 & 2761.12 (73.68) & 6.4 hours & 3 & 2730.95 (58.04) & 7.2 hours & \cellcolor{teal!20}1 & \cellcolor{teal!20}3359.75 (710.07) \\
\textbf{\textsc{Gcc}} & 3 & 70.79 (2.48) & 1.7 hours & 2 & 70.17 (2.78) & 0.4 hours & \cellcolor{teal!20}1 & \cellcolor{teal!20}67.01 (2.93) \\
\textbf{\textsc{Clang}} & 2 & 68.91 (3.58) & 0.0 hours & 2 & 69.57 (2.20) & 0.2 hours & \cellcolor{teal!20}1 & \cellcolor{teal!20}68.65 (4.43) \\ \bottomrule
\end{tabular}

	\end{threeparttable}
    \end{adjustbox}
\end{table}

\subsection{RQ4: Sensitivity to $\alpha$}
\label{subsection:sensitivity_analysis}

\subsubsection{Method.} To study the sensitivity of \approach~to the $\alpha$---the probability of including diverse configurations under the fair imperfect-fidelity---in \textbf{RQ4}, we study five values: $\{0.1, 0.3, 0.5, 0.7, 0.9\}$. Scott Knott ESD is used to ensure statistical significance over 10 runs.


\subsubsection{Results.}
As shown in Figure~\ref{fig:alpha}, the performance of \approach~is sensitive to $\alpha$, but generally $\alpha=0.5$ achieves the best outcomes on both rank and average performance---neither too small nor too large $\alpha$ is ideal. This makes sense, because a too small $\alpha$ would cause too few diverse configurations to be archived, which may lose some configurations that perform well under the perfect-fidelity setting but are sub-optimal under the fair imperfect-fidelity setting. In contrast, when $\alpha$ is too large, too many diverse configurations might be archived, hence missing the chance to find those that perform well under both settings. Overall, we show that:




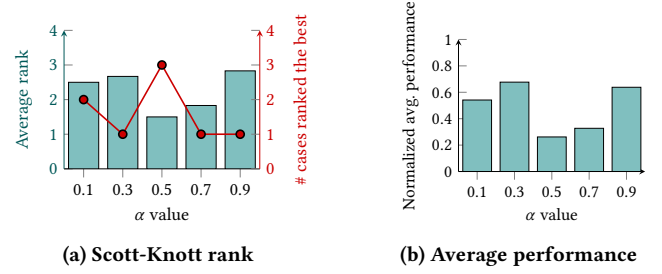
\begin{figure}[t!]
\centering
  \subfloat[Scott-Knott rank]{\begingroup
\pgfplotsset{compat=1.18}

\begin{tikzpicture}

\begin{axis}[
    name=leftaxis,
    height=4cm, width=5cm,
    ybar=0pt,
    bar width=15pt,
    xmin=0.5, xmax=5.5,
    ymin=0, ymax=4,
    xlabel={$\alpha$ value},
    xtick={1,2,3,4,5},
    xticklabels={0.1,0.3,0.5,0.7,0.9},
    ylabel={},
    axis y line*=left,
    axis x line*=bottom,
    y axis line style={-stealth,teal!70!black},
    yticklabel style={teal!70!black},
    legend cell align={left},
    legend style={draw=none, fill=none, at={(0.02,0.98)}, anchor=north west},
]
\addplot[fill=teal!50, draw=black] coordinates {
    (1,2.5)
    (2,2.67)
    (3,1.5)
    (4,1.83)
    (5,2.83)
};
\end{axis}

\begin{axis}[
    name=rightaxis,
    height=4cm, width=5cm,
    at={(leftaxis.south west)}, anchor=south west,
    xmin=0.5, xmax=5.5,
    ymin=0, ymax=4,
    axis y line*=right,
    axis x line=none,
    y axis line style={-stealth,red!80!black},
    yticklabel style={red!80!black},
    ylabel={},
    legend cell align={left},
    legend style={draw=none, fill=none, at={(0.98,0.98)}, anchor=north east},
]
\addplot+[
    mark=*,
    thick,
    draw=red!80!black,
    mark options={fill=red!80!black,draw=black}
] coordinates {
    (1,2)
    (2,1)
    (3,3)
    (4,1)
    (5,1)
};
\end{axis}

\node[rotate=90, text=teal!70!black, anchor=center]
    at ($(leftaxis.west)+(-20pt,0)$) {Average rank};
\node[rotate=90, text=red!80!black, anchor=center]
    at ($(rightaxis.east)+(20pt,0)$) {$\#$ cases ranked the best};

\end{tikzpicture}

\endgroup}
   \label{fig:alpha-rank}
\hfill
  \subfloat[Average performance]{\begingroup
\pgfplotsset{compat=1.18} 

\begin{tikzpicture}

\begin{axis}[
    name=leftaxis,
    height=4cm, width=5cm,
    ybar=0pt,
    bar width=15pt,
    xmin=0.5, xmax=5.5,
    ymin=0, ymax=1,
    xlabel={$\alpha$ value},
    xtick={1,2,3,4,5},
    xticklabels={0.1,0.3,0.5,0.7,0.9},
    ylabel={},                         
    axis y line*=left,
    axis x line*=bottom,
    x axis line style={-stealth},
    y axis line style={-stealth},
    legend cell align={left},
    legend style={draw=none, fill=none, at={(0.02,0.98)}, anchor=north west},
]
\addplot[fill=teal!50, draw=black] coordinates {
  (1,0.5415) (2, 0.6768) (3,0.2618) (4,0.3273) (5,0.638)
};
\end{axis}

\node[rotate=90, anchor=center] at ($(leftaxis.west)+(-25pt,0)$) {Normalized avg. performance};

\end{tikzpicture}
\endgroup}
    \label{fig:alpha-perf}
\caption{The sensitivity of \approach~to $\alpha$ values over all systems/runs (smaller normalized performance is preferred).}
\label{fig:alpha}
\end{figure}

\begin{quotebox}
   \noindent
   \textit{\textbf{RQ4:} \approach~is sensitives to  $\alpha$, but $\alpha=0.5$ is generally the best.}
\end{quotebox}

\section{Discussion}
\label{section:discussion}

\subsection{How Imperfect-Fidelity Discovery Helps?} 
\begin{figure}[t]
    \centering
    \subfloat[Iteration $k+1$]{\begingroup
\pgfplotsset{compat=1.18}

\begin{tikzpicture}
\begin{axis}[
    width=4cm,
    height=4cm,
    xmin=0.60, xmax=1.0,
    ymin=40,  ymax=75,
    xlabel={Fidelity perfection},
    ylabel={Cost (s)},
    xlabel style={
    font=\scriptsize,
    at={(axis description cs:0.5,-0.15)},  
    anchor=north
    },
    ylabel style={
        font=\scriptsize,
        at={(axis description cs:-0.2,0.5)},  
        anchor=south
    },
    xlabel style={font=\footnotesize},
    ylabel style={font=\footnotesize},
    tick label style={font=\footnotesize},
    axis x line*=bottom,
    axis y line*=left,
    x axis line style={-stealth},
    y axis line style={-stealth},
    enlargelimits=false,
    clip=false,
]

\addplot[
    only marks,
    mark=o,
    mark size=3.2pt,
    line width=0.8pt,
    draw=teal,
    fill=none,
]
table[row sep=\\]{
0.7576 52.3242\\
0.9273 57.7272\\
0.9515 61.1346\\
};

\addplot[
    only marks,
    mark=star,
    mark size=2.3pt,
    draw=red,
    fill=red,
]
table[row sep=\\]{
0.7576 52.3242\\
0.9273 57.7272\\
0.9515 61.1346\\
0.7333 56.9884\\
0.9152 58.4099\\
0.8909 70.3802\\
0.8545 59.9348\\
0.8667 68.2702\\
0.6848 60.9927\\
0.8424 69.0912\\
};





\end{axis}
\end{tikzpicture}

\endgroup}
    \hspace{-0.2cm}
      \subfloat[Iteration $k+2$]{\begingroup
\pgfplotsset{compat=1.18}

\begin{tikzpicture}
\begin{axis}[
    width=4cm,
    height=4cm,
    xmin=0.60, xmax=1.0,
    ymin=40,  ymax=75,
    xlabel={Fidelity perfection},
    ylabel={Cost (s)},
    xlabel style={
    font=\scriptsize,
    at={(axis description cs:0.5,-0.15)},  
    anchor=north
    },
    ylabel style={
        font=\scriptsize,
        at={(axis description cs:-0.2,0.5)},  
        anchor=south
    },
    xlabel style={font=\footnotesize},
    ylabel style={font=\footnotesize},
    tick label style={font=\footnotesize},
    axis x line*=bottom,
    axis y line*=left,
    x axis line style={-stealth},
    y axis line style={-stealth},
    enlargelimits=false,
    clip=false,
]

\addplot[
    only marks,
    mark=o,
    mark size=3.2pt,
    line width=0.8pt,
    draw=teal,
    fill=none,
]
table[row sep=\\]{
0.9515	61.1346\\
0.8667	45.1452\\
0.7697	43.5403\\
0.9394	50.1083\\
};

\addplot[
    only marks,
    mark=star,
    mark size=2.3pt,
    draw=red,
    fill=red,
]
table[row sep=\\]{
0.9515	61.1346\\
0.8667	45.1452\\
0.7697	43.5403\\
0.9394	50.1083\\
0.9394	73.2469\\
0.7576	52.3242\\
0.9273	57.7272\\
0.8061	54.3945\\
0.9152	58.4099\\
0.7333	56.9884\\
};




\end{axis}
\end{tikzpicture}

\endgroup}
      \hspace{-0.2cm}
      \subfloat[Iteration $k+3$]{\begingroup
\pgfplotsset{compat=1.18}

\begin{tikzpicture}
\begin{axis}[
    width=4cm,
    height=4cm,
    xmin=0.60, xmax=1.0,
    ymin=40,  ymax=75,
    xlabel={Fidelity perfection},
    ylabel={Cost (s)},
    xlabel style={
    font=\scriptsize,
    at={(axis description cs:0.5,-0.15)},  
    anchor=north
    },
    ylabel style={
        font=\scriptsize,
        at={(axis description cs:-0.2,0.5)},  
        anchor=south
    },
    xlabel style={font=\footnotesize},
    ylabel style={font=\footnotesize},
    tick label style={font=\footnotesize},
    axis x line*=bottom,
    axis y line*=left,
    x axis line style={-stealth},
    y axis line style={-stealth},
    enlargelimits=false,
    clip=false,
]

\addplot[
    only marks,
    mark=o,
    mark size=3.2pt,
    line width=0.8pt,
    draw=teal,
    fill=none,
]
table[row sep=\\]{
0.9515	61.1346\\
0.9394	50.1083\\
0.9152	41.4727\\
};

\addplot[
    only marks,
    mark=star,
    mark size=2.3pt,
    draw=red,
    fill=red,
]
table[row sep=\\]{
0.9515	61.1346\\
0.9394	50.1083\\
0.9152	41.4727\\
0.9394	73.2469\\
0.9273	57.7272\\
0.8667	45.1452\\
0.7697	43.5403\\
0.6364	44.7162\\
0.9152	62.3640\\
0.8424	50.6157\\
};




\node[
    align=center,
    inner sep=1pt,
    font=\scriptsize
] at (axis cs:0.78, 67) {fair imperfect-\\fidelity setting};

\draw[->, thin]
    (axis cs:0.78, 63) -- (axis cs:0.9,44);

\end{axis}
\end{tikzpicture}

\endgroup}

    \caption{The changes within imperfect-fidelity discovering on \textsc{MySQL}. ``\textcolor{red}{$\star$}'' and ``\textcolor{teal}{\small$\circ$}'' denote the imperfect-fidelity settings and the nondominated imperfect-fidelity settings, respectively, where the perfect-fidelity setting has a cost $\approx 250$s.}
    \label{fig:fidelity_evo}
\end{figure}
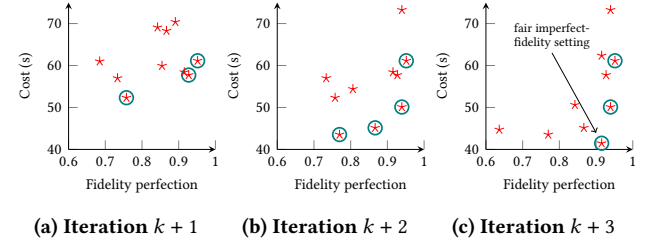

To understand how \approach~benefits from the imperfect-fidelity discovery, Figure~\ref{fig:fidelity_evo} illustrates how the explored imperfect-fidelity settings evolve across iterations when tuning \textsc{MySQL}. We can see that \approach~progressively reveals candidate imperfect-fidelity settings that yield better fidelity perfection at lower costs, enabling \approach~with a good start for both ``wideness'' and ``depth'' of tuning. 




\subsection{Why Tuning for Imperfect-Fidelity Work?} 
\begin{figure}[!t]
 \centering
    \begin{tikzpicture}
    \begin{axis}[
      axis x line  = bottom,
            axis y line  = left  ,
        width=10.5cm, height=6.7cm, 
        xlabel={Iteration in \approach},
        ylabel={Throughput (tps)},
        xmin=1, xmax=45,
        xtick={1,10,20,30,40},
        ymin=240, ymax=470,
        ytick={250,300,350,400,450,500,550,600},
            legend cell align=left,
          legend columns=1,
        legend style={
        cells={align=left},
          draw=none, fill=none,
          at={(1.05,0.35)}, anchor=south east,
          legend cell align=left
        }
    ]
\addplot[
red,
mark=o,mark options={thick,solid},
ultra thick, forget plot
] coordinates {
(1, 341.2900)
(4, 401.6055)
(7, 404.9410)
(10, 405.9460)
(13, 406.9795)
(16, 410.1465)
(19, 411.3655)
(22, 411.7435)
(25, 412.1720)
(28, 422.2715)
(31, 432.8225)
(34, 437.5390)
(37, 441.9480)
(40, 442.9220)
(44, 444.8515)
}; 

\addplot[
name path=upper_low,
draw=none, forget plot
] coordinates {
(1, 349.2900)
(4, 409.7622)
(7, 412.6567)
(10, 413.1545)
(13, 413.4818)
(16, 416.6537)
(19, 417.3477)
(22, 417.4981)
(25, 417.6814)
(28, 429.7550)
(31, 438.2520)
(34, 442.9842)
(37, 447.5704)
(40, 447.8590)
(44, 450.5851)

}; 

\addplot[
name path=lower_low,
draw=none, forget plot
] coordinates {
(1, 333.2900)
(4, 393.4488)
(7, 397.2253)
(10, 398.7375)
(13, 400.4772)
(16, 403.6393)
(19, 405.3833)
(22, 405.9889)
(25, 406.6626)
(28, 414.7880)
(31, 427.3930)
(34, 432.0938)
(37, 436.3256)
(40, 437.9850)
(44, 439.1179)
}; 

\addplot[
    red!70,
    fill opacity=0.30, forget plot
] fill between[
    of=upper_low and lower_low,
];

\addplot[
teal,
mark=square,mark options={thick},
ultra thick, forget plot
] coordinates {
(1, 257.8133)
(4, 258.4577)
(7, 260.6814)
(10, 260.6814)
(13, 260.6814)
(16, 260.6814)
(19, 260.6814)
(22, 260.6814)
(25, 288.3070)
(28, 293.0300)
(31, 301.7805)
(34, 307.7195)
(37, 312.4530)
(40, 312.4530)
(44, 353.9050)

}; 

\addplot[
name path=upper_hi,
draw=none, forget plot
] coordinates {
(1, 281.9686)
(4, 285.2355)
(7, 287.7021)
(10, 287.7021)
(13, 287.7021)
(16, 287.7021)
(19, 287.7021)
(22, 287.7021)
(25, 316.4572)
(28, 323.1328)
(31, 332.6957)
(34, 340.7989)
(37, 344.9272)
(40, 344.9272)
(44, 360.7290)

}; 

\addplot[
name path=lower_hi,
draw=none, forget plot
] coordinates {
(1, 233.6581)
(4, 231.6799)
(7, 233.6607)
(10, 233.6607)
(13, 233.6607)
(16, 233.6607)
(19, 233.6607)
(22, 233.6607)
(25, 260.1568)
(28, 262.9272)
(31, 270.8653)
(34, 274.6401)
(37, 279.9788)
(40, 279.9788)
(44, 347.0810)

}; 

\addplot[
teal!30,
fill opacity=0.30, forget plot
] fill between[
of=upper_hi and lower_hi,
];

\addplot[
    gray,
    dashed,
    very thick, forget plot
] coordinates {
    (25,10)
    (25,490)
};

\draw[<->, thick]
(axis cs:24, 450)
--
(axis cs:1.5, 450)
node[midway, above, yshift=2pt]
{\small Progressive Diversity Sampling};

\draw[<->, thick]
(axis cs:26, 450)  
--
(axis cs:44, 450) 
node[midway, above, yshift=2pt]
{\small Diversity-Preserved Tuning};

\addlegendimage{red, mark=o, mark options={thick,solid}, ultra thick}
\addlegendentry{Measuring under fair imperfect-fidelity setting \\(averaging all configurations in the sample set/population)}

\addlegendimage{teal, mark=square, mark options={thick}, ultra thick}
\addlegendentry{Measuring under perfect-fidelity setting \\(averaging the top $n$ configurations in the archive $\boldsymbol{\mathbfcal{A}}$)}

\end{axis}

\end{tikzpicture}
    \caption{Tuning trajectory of \textsc{MySQL} under the fair imperfect-fidelity and testing the corresponding configurations found thereof under the perfect-fidelity setting.}
   \label{fig:diss}
\end{figure}
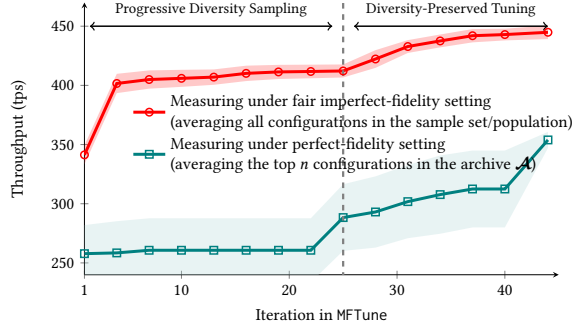

To disclose why tuning under a fair imperfect-fidelity setting $\boldsymbol{z}_{fair}$ works, Figure~\ref{fig:diss} plots an example of the tuning trajectory under the fair imperfect-fidelity setting $\boldsymbol{z}_{fair}$ and the changing top $n$ configurations in the archive $\mathbfcal{A}$ when tested under the perfect-fidelity setting $\boldsymbol{z}^{*}$. We see that the progressive diversity sampling increases the performance measured under both $\boldsymbol{z}_{fair}$ and $\boldsymbol{z}^{*}$ at a slow but steady pace. At around the $25$th iteration, the large improvement of the archive under $\boldsymbol{z}^{*}$ is due to the top configurations under  $\boldsymbol{z}_{fair}$ being pushed in. Both trace increases more steeply afterwards, especially when the overall top-performing configurations for $\boldsymbol{z}_{fair}$ are pushed to the archive again at the $40$th iteration.

Notably, the wall-clock time required for tuning/exploring the configuration under $\boldsymbol{z}_{fair}$ is only $\approx10$ hours, compared with the $\approx55$ hours if we were to tune/explore the same configurations under $\boldsymbol{z}^*$. This means that, if given the same wall-clock time (e.g., $10$ hours), the number of configurations that can be explored under $\boldsymbol{z}_{fair}$ would be much larger than that under $\boldsymbol{z}^*$ ($782$ vs. $142$), leading to superior results measured under $\boldsymbol{z}^*$ even though $\boldsymbol{z}_{fair}$ is imperfect.


The above evidence supports the rationale of \approach's superiority: finding the right imperfect-fidelity setting and properly tuning configurations can greatly help the tuning under the perfect-fidelity setting---a reflection of our hypothesis mentioned in \S\ref{section:introduction}.

\subsection{Threats to Validity}
\label{section:threats_to_validity}

\revision{\textbf{Internal threats} to validity could be raised from the parameter setting of $\alpha$, which controls the probability of triggering full-fidelity evaluations during low-fidelity guided tuning. In our study, we set $\alpha = 0.5$, following empirical evidence from our sensitivity analysis (\textbf{RQ3}), where this value offered a favorable balance between exploration cost and evaluation reliability. Similarly, parameter $l$ controls the number of paired configurations for estimating fidelity perfection. We set $l=10$ to strike a trade-off between estimation reliability and discovery cost: a large $l$ may provide a more reliable estimate but also incurs higher cost for fidelity discovery. Nevertheless, we recognize that the best value of these parameters may vary across systems, and exploring alternative settings could further enhance robustness.}

\section{Related Work}
\label{section:related_work}

\textbf{Tuning with or without models.} Model-free tuners address configuration tuning by guiding the search process exclusively through direct system measurements without models~\cite{behzad2013taming,chen2018sampling,DBLP:journals/corr/abs-2501-00840}. For example, \texttt{GA} has inspired many tuners that work on a population of configurations~\cite{behzad2013taming, shahbazian2020equal}. In contrast, model-based tuners learn a surrogate model from measured configuration-performance, paired with other heuristics, to accelerate the tuning~\cite{gong2023predicting,gong2024predicting,van2017automatic,chen2021efficient,zhu2023compiler,chen2018tvm,DBLP:conf/ijcai/ChenCDS4RAG}. However, all of the above are single-fidelity tuners, where each configuration is measured at the perfect-fidelity setting, guaranteeing accurate measurements at the expense of high costs. 
\approach, in contrast, explicitly leverages the imperfect-fidelity setting to improve the tuning and budget utilization.








\textbf{Multi-workloads tuners.} Recent works have transferred knowledge and reused information from past tuning tasks/environments/workloads, which might be similar to the fidelity: 

\begin{itemize}
    \item \textbf{Workload mapping} (e.g., \texttt{OtterTune}~\cite{van2017automatic}) identifies the most similar historical environment/workload and reuses the data to build a surrogate model.
    \item \textbf{Model ensemble} (e.g., \texttt{ResTune}~\cite{zhang2021restune}) combines multiple surrogate models trained on prior environments/workloads and generalizes such to a new workload for tuning.
    \item \textbf{Option pruning} (e.g.,  \texttt{OpAdviser}~\cite{zhang2023efficient}) learns prior data under certain environments/workloads to identify key options/ranges for tuning under a new workload.
\end{itemize}




\approach~fundamentally differs from them in several aspects:

\begin{itemize}
    \item They assume a reactive situation where a tuner can only reuse historically explored and structurally similar environments/workloads, limiting their ability to adapt to a completely unforeseen environment; while \approach~is a proactive tuner that actively explores unknown fidelity settings.
    \item They assume dozens of environments/workloads; \approach\ scales to ten thousand environments/fidelity settings.
    \item They typically reuse the knowledge of other environments/workloads for updating the surrogate model or dimensionality reduction, while \approach~leverages them for consolidating the tuning process directly.
    
\end{itemize}



\textbf{Multi-fidelity optimization.} Multiple fidelity settings have been considered primarily in the context of HPO~\cite{hu2019multi,forrester2007multi,klein2017fast}. A representative solution is \texttt{Hyperband}~\cite{li2018hyperband}, which samples a set of random solutions and progressively allocates more resources to promising ones via successive halving~\cite{jamieson2016non}. Variants such as \texttt{BOHB}~\cite{falkner2018bohb} and \texttt{DEHB}~\cite{awad-ijcai21} extend \texttt{Hyperband} by incorporating Bayesian Optimization and Differential Evolution, respectively, to enhance optimization efficiency. However, as mentioned, they mostly consider a limited dimension of the fidelity factors and have also assumed a monotonic correlation between fidelity perfection and cost~\cite{DBLP:journals/corr/abs-2602-00788,echevarrieta2024speeding,carstensen2025frozen}, which, as we have shown, is not the case for configurable systems. This work formulates a conceptual framework specifically for system configuration tuning while presenting \approach~as the tuner that works effectively without the above constraints/assumptions.

\section{Conclusion}
\label{section:conclusion}

This paper presents \approach, a tuner that tackles system configuration tuning by proactively exploring and exploiting the imperfect-fidelity setting. Drawing on a newly formulated conceptual framework of multi-fidelity for configurable systems, \approach~explore in the scale of more than ten thousand imperfect-fidelity settings, finding the most fair one for tuning, which then seeds the tuning under the perfect-fidelity setting for improved budget utilization and superior results. Experiments against 10 state-of-the-art tuners and under six real-world systems show that \approach~achieves better results on $83.33$\% cases with up to $19.34$\% improvements, while doing so by generally saving hours' budget. Looking forward, we envision that this work can inspire new research directions that integrate multi-fidelity with diverse optimization paradigms, paving the way for more advanced configuration tuning for configurable systems. 




\begin{acks}
    This work was supported by an NSFC Grant (62372084) and a UKRI Grant (10054084).
\end{acks}

\section*{Data Availability Statement}
All source code and data are available at: \textcolor{blue}{\texttt{\url{https://github.com/ideas-labo/mftune}}} and \textcolor{blue}{\texttt{\url{https://zenodo.org/records/21364961}}}.

\balance
\bibliographystyle{ACM-Reference-Format}
\bibliography{reference}

@article{gong2024dividable,
  author       = {Jingzhi Gong and
                  Tao Chen and
                  Rami Bahsoon},
  title        = {Dividable Configuration Performance Learning},
  journal      = {{IEEE} Trans. Software Eng.},
  volume       = {51},
  number       = {1},
  pages        = {106--134},
  year         = {2025},
  url          = {https://doi.org/10.1109/TSE.2024.3491945},
  doi          = {10.1109/TSE.2024.3491945},
  bibsource    = {dblp computer science bibliography, https://dblp.org}
}

@inproceedings{DBLP:conf/icse/LiangHC25,
  author       = {Hongyuan Liang and
                  Yue Huang and
                  Tao Chen},
  title        = {The Same Only Different: On Information Modality for Configuration
                  Performance Analysis},
  booktitle    = {47th {IEEE/ACM} International Conference on Software Engineering,
                  {ICSE} 2025, Ottawa, ON, Canada, April 26 - May 6, 2025},
  pages        = {2522--2534},
  publisher    = {{IEEE}},
  year         = {2025},
  url          = {https://doi.org/10.1109/ICSE55347.2025.00212},
  doi          = {10.1109/ICSE55347.2025.00212},
  bibsource    = {dblp computer science bibliography, https://dblp.org}
}

@inproceedings{DBLP:conf/kbse/XiongC25,
  author       = {Gangda Xiong and
                  Tao Chen},
  title        = {CoTune: Co-evolutionary Configuration Tuning},
  booktitle    = {40th {IEEE/ACM} International Conference on Automated Software Engineering,
                  {ASE} 2025, Seoul, Korea, Republic of, November 16-20, 2025},
  pages        = {1490--1502},
  publisher    = {{IEEE}},
  year         = {2025},
  url          = {https://doi.org/10.1109/ASE63991.2025.00126},
  doi          = {10.1109/ASE63991.2025.00126},
  bibsource    = {dblp computer science bibliography, https://dblp.org}
}

@article{10.1145/3803859,
author = {Ye, Yulong and Liang, Hongyuan and Jiang, Chao and Li, Miqing and Chen, Tao},
title = {Revealing Domain-Spatiality Patterns for Configuration Tuning: Domain Knowledge Meets Fitness Landscapes},
year = {2026},
publisher = {Association for Computing Machinery},
address = {New York, NY, USA},
issn = {1049-331X},
url = {https://doi.org/10.1145/3803859},
doi = {10.1145/3803859},
note = {Just Accepted},
journal = {ACM Trans. Softw. Eng. Methodol.},
month = mar
}

@inproceedings{DBLP:conf/ijcai/ChenCDS4RAG,
  author       = {Pengzhou Chen and
                  Tao Chen},
  title        = {CDS4RAG: Cyclic Dual-Sequential Hyperparameter Optimization for RAG},
  booktitle    = {Proceedings of the 35th International Joint Conference on Artificial
                  Intelligence, {IJCAI} 2026, Bremen, Germany, 15-21
                  August 2026},
  publisher    = {ijcai.org},
  year         = {2026}
}

@article{DBLP:journals/corr/abs-2602-00788,
  author       = {Md. Abir Hossen and
                  Mohammad Ali Javidian and
                  Vignesh Narayanan and
                  Jason M. O'Kane and
                  Pooyan Jamshidi},
  title        = {Multi-Objective Multi-Fidelity Bayesian Optimization with Causal Priors},
  journal      = {CoRR},
  volume       = {abs/2602.00788},
  year         = {2026},
  url          = {https://doi.org/10.48550/arXiv.2602.00788},
  doi          = {10.48550/ARXIV.2602.00788},
  eprinttype   = {arXiv},
  eprint       = {2602.00788},
  bibsource    = {dblp computer science bibliography, https://dblp.org}
}

@inproceedings{chen2018tvm,
  author       = {Tianqi Chen and
                  Thierry Moreau and
                  Ziheng Jiang and
                  Lianmin Zheng and
                  Eddie Q. Yan and
                  Haichen Shen and
                  Meghan Cowan and
                  Leyuan Wang and
                  Yuwei Hu and
                  Luis Ceze and
                  Carlos Guestrin and
                  Arvind Krishnamurthy},
  editor       = {Andrea C. Arpaci{-}Dusseau and
                  Geoff Voelker},
  title        = {{TVM:} An Automated End-to-End Optimizing Compiler for Deep Learning},
  booktitle    = {13th {USENIX} Symposium on Operating Systems Design and Implementation,
                  {OSDI} 2018, Carlsbad, CA, USA, October 8-10, 2018},
  pages        = {578--594},
  publisher    = {{USENIX} Association},
  year         = {2018},
  url          = {https://www.usenix.org/conference/osdi18/presentation/chen},
  bibsource    = {dblp computer science bibliography, https://dblp.org}
}

@inproceedings{hu2023hydro,
  author       = {Qinghao Hu and
                  Zhisheng Ye and
                  Meng Zhang and
                  Qiaoling Chen and
                  Peng Sun and
                  Yonggang Wen and
                  Tianwei Zhang},
  editor       = {Roxana Geambasu and
                  Ed Nightingale},
  title        = {Hydro: Surrogate-Based Hyperparameter Tuning Service in Datacenters},
  booktitle    = {17th {USENIX} Symposium on Operating Systems Design and Implementation,
                  {OSDI} 2023, Boston, MA, USA, July 10-12, 2023},
  pages        = {757--777},
  publisher    = {{USENIX} Association},
  year         = {2023},
  url          = {https://www.usenix.org/conference/osdi23/presentation/hu},
  bibsource    = {dblp computer science bibliography, https://dblp.org}
}

@article{zhang2023efficient,
  author       = {Xinyi Zhang and
                  Hong Wu and
                  Yang Li and
                  Zhengju Tang and
                  Jian Tan and
                  Feifei Li and
                  Bin Cui},
  title        = {An Efficient Transfer Learning Based Configuration Adviser for Database
                  Tuning},
  journal      = {Proc. {VLDB} Endow.},
  volume       = {17},
  number       = {3},
  pages        = {539--552},
  year         = {2023},
  url          = {https://www.vldb.org/pvldb/vol17/p539-zhang.pdf},
  doi          = {10.14778/3632093.3632114},
  bibsource    = {dblp computer science bibliography, https://dblp.org}
}

@misc{clang_v,
  year = 2024,
  title        = {Clang},
  howpublished = {\url{https://releases.llvm.org/download.html}},
  note         = {Version: 17.0.6. Accessed: 2024-02-20}
}

@misc{gcc_v,
  year = 2024,
  title        = {Gcc},
  howpublished = {\url{https://gcc.gnu.org/pub/gcc/releases/gcc-14.2.0/}},
  note         = {Version: 14.2. Accessed: 2024-02-20}
}

@misc{httpd_v,
  year = 2024,
  title        = {Httpd},
  howpublished = {\url{https://httpd.apache.org/download.cgi}},
  note         = {Version: 2.4.57. Accessed: 2024-02-20}
}

@misc{tomcat_v,
  year = 2024,
  title        = {Tomcat},
  howpublished = {\url{https://tomcat.apache.org/download-10.cgi}},
  note         = {Version: 10.1.34. Accessed: 2024-02-20}
}

@misc{postgresql_v,
  year = 2024,
  title        = {PostgreSQL},
  howpublished = {\url{https://www.postgresql.org/docs/release/12.7/}},
  note         = {Version: 12.7. Accessed: 2024-02-20}
}

@misc{mysql_v,
  year = 2024,
  title        = {MySQL},
  howpublished = {\url{https://dev.mysql.com/doc/relnotes/mysql/5.7/en/}},
  note         = {Version: 5.7.19. Accessed: 2024-02-20}
}

@misc{sysbench,
  year = 2024,
  title        = {Sysbench},
  howpublished = {\url{https://github.com/akopytov/sysbench}},
  note         = {Accessed: 2024-02-20}
}

@misc{Wrk,
  year = 2024,
  title        = {Wrk},
  howpublished = {\url{https://github.com/wg/wrk}},
  note         = {Accessed: 2024-02-20}
}

@misc{Csmith,
  year = 2024,
  title        = {Csmith},
  howpublished = {\url{https://github.com/csmith-project/csmith}},
  note         = {Accessed: 2024-02-20}
}

@article{hauke2011comparison,
  title={Comparison of values of Pearson's and Spearman's correlation coefficients on the same sets of data},
  author={Hauke, Jan and Kossowski, Tomasz},
  journal={Quaestiones geographicae},
  volume={30},
  number={2},
  pages={87--93},
  year={2011}
}

@article{mchugh2011multiple,
  title={Multiple comparison analysis testing in ANOVA},
  author={McHugh, Mary L},
  journal={Biochemia medica},
  volume={21},
  number={3},
  pages={203--209},
  year={2011},
  publisher={Hrvatsko dru{\v{s}}tvo za medicinsku biokemiju i laboratorijsku medicinu}
}

@article{echevarrieta2024speeding,
  author       = {Judith Echevarrieta and
                  Etor Arza and
                  Aritz P{\'{e}}rez},
  title        = {Speeding-Up Evolutionary Algorithms to Solve Black-Box Optimization
                  Problems},
  journal      = {{IEEE} Trans. Evol. Comput.},
  volume       = {29},
  number       = {1},
  pages        = {117--131},
  year         = {2025},
  url          = {https://doi.org/10.1109/TEVC.2024.3352450},
  doi          = {10.1109/TEVC.2024.3352450},
  bibsource    = {dblp computer science bibliography, https://dblp.org}
}

@ARTICLE{deb2002fast,
  author={Deb, K. and Pratap, A. and Agarwal, S. and Meyarivan, T.},
  journal={{IEEE} Trans. Evol. Comput.}, 
  title={A fast and elitist multiobjective genetic algorithm: NSGA-II}, 
  year={2002},
  volume={6},
  number={2},
  pages={182-197},
  doi={10.1109/4235.996017}}

@inproceedings{mckay1992latin,
  author       = {Michael D. McKay},
  editor       = {Robert C. Crain},
  title        = {Latin Hypercube Sampling as a Tool in Uncertainty Analysis of Computer
                  Models},
  booktitle    = {Proceedings of the 24th Winter Simulation Conference, Arlington, VA,
                  USA, December 13-16, 1992},
  pages        = {557--564},
  publisher    = {{ACM} Press},
  year         = {1992},
  url          = {https://doi.org/10.1145/167293.167637},
  doi          = {10.1145/167293.167637},
  bibsource    = {dblp computer science bibliography, https://dblp.org}
}

@inproceedings{DBLP:conf/nips/EggenspergerMMF21,
  author       = {Katharina Eggensperger and
                  Philipp M{\"{u}}ller and
                  Neeratyoy Mallik and
                  Matthias Feurer and
                  Ren{\'{e}} Sass and
                  Aaron Klein and
                  Noor H. Awad and
                  Marius Lindauer and
                  Frank Hutter},
  editor       = {Joaquin Vanschoren and
                  Sai{-}Kit Yeung},
  title        = {HPOBench: {A} Collection of Reproducible Multi-Fidelity Benchmark
                  Problems for {HPO}},
  booktitle    = {Proceedings of the Neural Information Processing Systems Track on
                  Datasets and Benchmarks 1, NeurIPS Datasets and Benchmarks 2021, December
                  2021, virtual},
  year         = {2021},
  url          = {https://datasets-benchmarks-proceedings.neurips.cc/paper/2021/hash/93db85ed909c13838ff95ccfa94cebd9-Abstract-round2.html},
  bibsource    = {dblp computer science bibliography, https://dblp.org}
}

@InProceedings{carstensen2025frozen,
  title = 	 {Frozen Layers: Memory-efficient Many-fidelity Hyperparameter Optimization},
  author =       {Carstensen, Timur and Mallik, Neeratyoy and Hutter, Frank and Rapp, Martin},
  booktitle = 	 {Proceedings of the Fourth International Conference on Automated Machine Learning},
  pages = 	 {4/1--24},
  year = 	 {2025},
  editor = 	 {Akoglu, Leman and Doerr, Carola and van Rijn, Jan N. and Garnett, Roman and Gardner, Jacob R.},
  volume = 	 {293},
  series = 	 {Proceedings of Machine Learning Research},
  month = 	 {08--11 Sep},
  publisher =    {PMLR},
  url = 	 {https://proceedings.mlr.press/v293/carstensen25a.html}
}

@inproceedings{jamieson2016non,
  author       = {Kevin Jamieson and
                  Ameet Talwalkar},
  editor       = {Arthur Gretton and
                  Christian C. Robert},
  title        = {Non-stochastic Best Arm Identification and Hyperparameter Optimization},
  booktitle    = {Proceedings of the 19th International Conference on Artificial Intelligence
                  and Statistics, {AISTATS} 2016, Cadiz, Spain, May 9-11, 2016},
  series       = {{JMLR} Workshop and Conference Proceedings},
  volume       = {51},
  pages        = {240--248},
  publisher    = {JMLR.org},
  year         = {2016},
  url          = {http://proceedings.mlr.press/v51/jamieson16.html},
  bibsource    = {dblp computer science bibliography, https://dblp.org}
}

@inproceedings{kandasamy2017multi,
  author       = {Kirthevasan Kandasamy and
                  Gautam Dasarathy and
                  Jeff G. Schneider and
                  Barnab{\'{a}}s P{\'{o}}czos},
  editor       = {Doina Precup and
                  Yee Whye Teh},
  title        = {Multi-fidelity Bayesian Optimisation with Continuous Approximations},
  booktitle    = {Proceedings of the 34th International Conference on Machine Learning,
                  {ICML} 2017, Sydney, NSW, Australia, 6-11 August 2017},
  series       = {Proceedings of Machine Learning Research},
  volume       = {70},
  pages        = {1799--1808},
  publisher    = {{PMLR}},
  year         = {2017},
  url          = {http://proceedings.mlr.press/v70/kandasamy17a.html},
  bibsource    = {dblp computer science bibliography, https://dblp.org}
}

@article{forrester2007multi,
  title={Multi-fidelity optimization via surrogate modelling},
  author={Forrester, Alexander IJ and S{\'o}bester, Andr{\'a}s and Keane, Andy J},
  journal={Proceedings of the royal society a: mathematical, physical and engineering sciences},
  volume={463},
  number={2088},
  pages={3251--3269},
  year={2007},
  publisher={The Royal Society London}
}

@inproceedings{zhu2023compiler,
  author       = {Mingxuan Zhu and
                  Dan Hao},
  title        = {Compiler Auto-Tuning via Critical Flag Selection},
  booktitle    = {38th {IEEE/ACM} International Conference on Automated Software Engineering,
                  {ASE} 2023, Luxembourg, September 11-15, 2023},
  pages        = {1000--1011},
  publisher    = {{IEEE}},
  year         = {2023},
  url          = {https://doi.org/10.1109/ASE56229.2023.00209},
  doi          = {10.1109/ASE56229.2023.00209},
  bibsource    = {dblp computer science bibliography, https://dblp.org}
}

@inproceedings{van2017automatic,
  author       = {Dana Van Aken and
                  Andrew Pavlo and
                  Geoffrey J. Gordon and
                  Bohan Zhang},
  editor       = {Semih Salihoglu and
                  Wenchao Zhou and
                  Rada Chirkova and
                  Jun Yang and
                  Dan Suciu},
  title        = {Automatic Database Management System Tuning Through Large-scale Machine
                  Learning},
  booktitle    = {Proceedings of the 2017 {ACM} International Conference on Management
                  of Data, {SIGMOD} Conference 2017, Chicago, IL, USA, May 14-19, 2017},
  pages        = {1009--1024},
  publisher    = {{ACM}},
  year         = {2017},
  url          = {https://doi.org/10.1145/3035918.3064029},
  doi          = {10.1145/3035918.3064029},
  bibsource    = {dblp computer science bibliography, https://dblp.org}
}

@inproceedings{behzad2013taming,
  author       = {Babak Behzad and
                  Huong Vu Thanh Luu and
                  Joseph Huchette and
                  Surendra Byna and
                  Prabhat and
                  Ruth A. Aydt and
                  Quincey Koziol and
                  Marc Snir},
  editor       = {William Gropp and
                  Satoshi Matsuoka},
  title        = {Taming parallel {I/O} complexity with auto-tuning},
  booktitle    = {International Conference for High Performance Computing, Networking,
                  Storage and Analysis, SC'13, Denver, CO, {USA} - November 17 - 21,
                  2013},
  pages        = {68:1--68:12},
  publisher    = {{ACM}},
  year         = {2013},
  url          = {https://doi.org/10.1145/2503210.2503278},
  doi          = {10.1145/2503210.2503278},
  bibsource    = {dblp computer science bibliography, https://dblp.org}
}

@article{gong2024predicting,
  author       = {Jingzhi Gong and
                  Tao Chen},
  title        = {Predicting Configuration Performance in Multiple Environments with
                  Sequential Meta-Learning},
  journal      = {Proc. {ACM} Softw. Eng.},
  volume       = {1},
  number       = {{FSE}},
  pages        = {359--382},
  year         = {2024},
  url          = {https://doi.org/10.1145/3643743},
  doi          = {10.1145/3643743},
  bibsource    = {dblp computer science bibliography, https://dblp.org}
}

@inproceedings{gong2023predicting,
  author       = {Jingzhi Gong and
                  Tao Chen},
  editor       = {Satish Chandra and
                  Kelly Blincoe and
                  Paolo Tonella},
  title        = {Predicting Software Performance with Divide-and-Learn},
  booktitle    = {Proceedings of the 31st {ACM} Joint European Software Engineering
                  Conference and Symposium on the Foundations of Software Engineering,
                  {ESEC/FSE} 2023, San Francisco, CA, USA, December 3-9, 2023},
  pages        = {858--870},
  publisher    = {{ACM}},
  year         = {2023},
  url          = {https://doi.org/10.1145/3611643.3616334},
  doi          = {10.1145/3611643.3616334},
  bibsource    = {dblp computer science bibliography, https://dblp.org}
}

@article{tantithamthavorn2016empirical,
  author       = {Steffen Herbold},
  title        = {Comments on ScottKnottESD in Response to "An Empirical Comparison
                  of Model Validation Techniques for Defect Prediction Models"},
  journal      = {{IEEE} Trans. Software Eng.},
  volume       = {43},
  number       = {11},
  pages        = {1091--1094},
  year         = {2017},
  url          = {https://doi.org/10.1109/TSE.2017.2748129},
  doi          = {10.1109/TSE.2017.2748129},
  bibsource    = {dblp computer science bibliography, https://dblp.org}
}

@article{DBLP:journals/pvldb/KanellisDKMCV22,
  author       = {Konstantinos Kanellis and
                  Cong Ding and
                  Brian Kroth and
                  Andreas M{\"{u}}ller and
                  Carlo Curino and
                  Shivaram Venkataraman},
  title        = {LlamaTune: Sample-Efficient {DBMS} Configuration Tuning},
  journal      = {Proc. {VLDB} Endow.},
  volume       = {15},
  number       = {11},
  pages        = {2953--2965},
  year         = {2022},
  url          = {https://www.vldb.org/pvldb/vol15/p2953-kanellis.pdf},
  doi          = {10.14778/3551793.3551844},
  bibsource    = {dblp computer science bibliography, https://dblp.org}
}

@inproceedings{shahbazian2020equal,
  author       = {Arman Shahbazian and
                  Suhrid Karthik and
                  Yuriy Brun and
                  Nenad Medvidovic},
  editor       = {Prem Devanbu and
                  Myra B. Cohen and
                  Thomas Zimmermann},
  title        = {eQual: informing early design decisions},
  booktitle    = {{ESEC/FSE} '20: 28th {ACM} Joint European Software Engineering Conference
                  and Symposium on the Foundations of Software Engineering, Virtual
                  Event, USA, November 8-13, 2020},
  pages        = {1039--1051},
  publisher    = {{ACM}},
  year         = {2020},
  url          = {https://doi.org/10.1145/3368089.3409749},
  doi          = {10.1145/3368089.3409749},
  bibsource    = {dblp computer science bibliography, https://dblp.org}
}

@inproceedings{mallik2023priorband,
  author       = {Neeratyoy Mallik and
                  Edward Bergman and
                  Carl Hvarfner and
                  Danny Stoll and
                  Maciej Janowski and
                  Marius Lindauer and
                  Luigi Nardi and
                  Frank Hutter},
  editor       = {Alice Oh and
                  Tristan Naumann and
                  Amir Globerson and
                  Kate Saenko and
                  Moritz Hardt and
                  Sergey Levine},
  title        = {PriorBand: Practical Hyperparameter Optimization in the Age of Deep
                  Learning},
  booktitle    = {Advances in Neural Information Processing Systems 36: Annual Conference
                  on Neural Information Processing Systems 2023, NeurIPS 2023, New Orleans,
                  LA, USA, December 10 - 16, 2023},
  year         = {2023},
  url          = {http://papers.nips.cc/paper\_files/paper/2023/hash/1704fe7aaff33a54802b83a016050ab8-Abstract-Conference.html},
  bibsource    = {dblp computer science bibliography, https://dblp.org}
}

@inproceedings{chen2025promisetune,
  title={PromiseTune: Unveiling Causally Promising and Explainable Configuration Tuning},
  author={Chen, Pengzhou and Chen, Tao},
  booktitle={2026 IEEE/ACM 48th International Conference on Software Engineering (ICSE)},
  year={2026}
}

@inproceedings{xiang2025dually,
  title={Dually Hierarchical Drift Adaptation for Online Configuration Performance Learning},
  author={Xiang, Zezhen and Gong, Jingzhi and Chen, Tao},
  booktitle={2026 IEEE/ACM 48th International Conference on Software Engineering (ICSE)},
  year={2026}
}

@article{cowen2022hebo,
  title={Hebo: Pushing the limits of sample-efficient hyper-parameter optimisation},
  author={Cowen-Rivers, Alexander I and Lyu, Wenlong and Tutunov, Rasul and Wang, Zhi and Grosnit, Antoine and Griffiths, Ryan Rhys and Maraval, Alexandre Max and Jianye, Hao and Wang, Jun and Peters, Jan and others},
  journal={Journal of Artificial Intelligence Research},
  volume={74},
  pages={1269--1349},
  year={2022}
}

@inproceedings{zhu2017bestconfig,
  author       = {Yuqing Zhu and
                  Jianxun Liu and
                  Mengying Guo and
                  Yungang Bao and
                  Wenlong Ma and
                  Zhuoyue Liu and
                  Kunpeng Song and
                  Yingchun Yang},
  title        = {BestConfig: tapping the performance potential of systems via automatic
                  configuration tuning},
  booktitle    = {Proceedings of the 2017 Symposium on Cloud Computing, SoCC 2017, Santa
                  Clara, CA, USA, September 24-27, 2017},
  pages        = {338--350},
  publisher    = {{ACM}},
  year         = {2017},
  url          = {https://doi.org/10.1145/3127479.3128605},
  doi          = {10.1145/3127479.3128605},
  bibsource    = {dblp computer science bibliography, https://dblp.org}
}

@inproceedings{snoek2012practical,
  author       = {Jasper Snoek and
                  Hugo Larochelle and
                  Ryan P. Adams},
  editor       = {Peter L. Bartlett and
                  Fernando C. N. Pereira and
                  Christopher J. C. Burges and
                  L{\'{e}}on Bottou and
                  Kilian Q. Weinberger},
  title        = {Practical Bayesian Optimization of Machine Learning Algorithms},
  booktitle    = {Advances in Neural Information Processing Systems 25: 26th Annual
                  Conference on Neural Information Processing Systems 2012. Proceedings
                  of a meeting held December 3-6, 2012, Lake Tahoe, Nevada, United States},
  pages        = {2960--2968},
  year         = {2012},
  url          = {https://proceedings.neurips.cc/paper/2012/hash/05311655a15b75fab86956663e1819cd-Abstract.html},
  bibsource    = {dblp computer science bibliography, https://dblp.org}
}

@inproceedings{awad-ijcai21,
  author       = {Noor H. Awad and
                  Neeratyoy Mallik and
                  Frank Hutter},
  editor       = {Zhi{-}Hua Zhou},
  title        = {{DEHB:} Evolutionary Hyberband for Scalable, Robust and Efficient
                  Hyperparameter Optimization},
  booktitle    = {Proceedings of the Thirtieth International Joint Conference on Artificial
                  Intelligence, {IJCAI} 2021, Virtual Event / Montreal, Canada, 19-27
                  August 2021},
  pages        = {2147--2153},
  publisher    = {ijcai.org},
  year         = {2021},
  url          = {https://doi.org/10.24963/ijcai.2021/296},
  doi          = {10.24963/IJCAI.2021/296},
  bibsource    = {dblp computer science bibliography, https://dblp.org}
}

@inproceedings{falkner2018bohb,
  author       = {Stefan Falkner and
                  Aaron Klein and
                  Frank Hutter},
  editor       = {Jennifer G. Dy and
                  Andreas Krause},
  title        = {{BOHB:} Robust and Efficient Hyperparameter Optimization at Scale},
  booktitle    = {Proceedings of the 35th International Conference on Machine Learning,
                  {ICML} 2018, Stockholmsm{\"{a}}ssan, Stockholm, Sweden, July
                  10-15, 2018},
  series       = {Proceedings of Machine Learning Research},
  volume       = {80},
  pages        = {1436--1445},
  publisher    = {{PMLR}},
  year         = {2018},
  url          = {http://proceedings.mlr.press/v80/falkner18a.html},
  bibsource    = {dblp computer science bibliography, https://dblp.org}
}

@article{li2018hyperband,
  author       = {Lisha Li and
                  Kevin Jamieson and
                  Giulia DeSalvo and
                  Afshin Rostamizadeh and
                  Ameet Talwalkar},
  title        = {Hyperband: {A} Novel Bandit-Based Approach to Hyperparameter Optimization},
  journal      = {J. Mach. Learn. Res.},
  volume       = {18},
  pages        = {185:1--185:52},
  year         = {2017},
  url          = {https://jmlr.org/papers/v18/16-558.html},
  bibsource    = {dblp computer science bibliography, https://dblp.org}
}

@inproceedings{domhan2015speeding,
  author       = {Tobias Domhan and
                  Jost Tobias Springenberg and
                  Frank Hutter},
  editor       = {Qiang Yang and
                  Michael J. Wooldridge},
  title        = {Speeding Up Automatic Hyperparameter Optimization of Deep Neural Networks
                  by Extrapolation of Learning Curves},
  booktitle    = {Proceedings of the Twenty-Fourth International Joint Conference on
                  Artificial Intelligence, {IJCAI} 2015, Buenos Aires, Argentina, July
                  25-31, 2015},
  pages        = {3460--3468},
  publisher    = {{AAAI} Press},
  year         = {2015},
  url          = {http://ijcai.org/Abstract/15/487},
  bibsource    = {dblp computer science bibliography, https://dblp.org}
}

@inproceedings{klein2017fast,
  author       = {Aaron Klein and
                  Stefan Falkner and
                  Simon Bartels and
                  Philipp Hennig and
                  Frank Hutter},
  editor       = {Aarti Singh and
                  Xiaojin (Jerry) Zhu},
  title        = {Fast Bayesian Optimization of Machine Learning Hyperparameters on
                  Large Datasets},
  booktitle    = {Proceedings of the 20th International Conference on Artificial Intelligence
                  and Statistics, {AISTATS} 2017, 20-22 April 2017, Fort Lauderdale,
                  FL, {USA}},
  series       = {Proceedings of Machine Learning Research},
  volume       = {54},
  pages        = {528--536},
  publisher    = {{PMLR}},
  year         = {2017},
  url          = {http://proceedings.mlr.press/v54/klein17a.html},
  bibsource    = {dblp computer science bibliography, https://dblp.org}
}

@inproceedings{hu2019multi,
  author       = {Yi{-}Qi Hu and
                  Yang Yu and
                  Wei{-}Wei Tu and
                  Qiang Yang and
                  Yuqiang Chen and
                  Wenyuan Dai},
  title        = {Multi-Fidelity Automatic Hyper-Parameter Tuning via Transfer Series
                  Expansion},
  booktitle    = {The Thirty-Third {AAAI} Conference on Artificial Intelligence, {AAAI}
                  2019, The Thirty-First Innovative Applications of Artificial Intelligence
                  Conference, {IAAI} 2019, The Ninth {AAAI} Symposium on Educational
                  Advances in Artificial Intelligence, {EAAI} 2019, Honolulu, Hawaii,
                  USA, January 27 - February 1, 2019},
  pages        = {3846--3853},
  publisher    = {{AAAI} Press},
  year         = {2019},
  url          = {https://doi.org/10.1609/aaai.v33i01.33013846},
  doi          = {10.1609/AAAI.V33I01.33013846},
  bibsource    = {dblp computer science bibliography, https://dblp.org}
}

@inproceedings{hutter2011sequential,
  author       = {Frank Hutter and
                  Holger H. Hoos and
                  Kevin Leyton{-}Brown},
  editor       = {Carlos A. Coello Coello},
  title        = {Sequential Model-Based Optimization for General Algorithm Configuration},
  booktitle    = {Learning and Intelligent Optimization - 5th International Conference,
                  {LION} 5, Rome, Italy, January 17-21, 2011. Selected Papers},
  series       = {Lecture Notes in Computer Science},
  volume       = {6683},
  pages        = {507--523},
  publisher    = {Springer},
  year         = {2011},
  url          = {https://doi.org/10.1007/978-3-642-25566-3\_40},
  doi          = {10.1007/978-3-642-25566-3\_40},
  bibsource    = {dblp computer science bibliography, https://dblp.org}
}

@inproceedings{chen2021efficient,
  author       = {Junjie Chen and
                  Ningxin Xu and
                  Peiqi Chen and
                  Hongyu Zhang},
  title        = {Efficient Compiler Autotuning via Bayesian Optimization},
  booktitle    = {43rd {IEEE/ACM} International Conference on Software Engineering,
                  {ICSE} 2021, Madrid, Spain, 22-30 May 2021},
  pages        = {1198--1209},
  publisher    = {{IEEE}},
  year         = {2021},
  url          = {https://doi.org/10.1109/ICSE43902.2021.00110},
  doi          = {10.1109/ICSE43902.2021.00110},
  bibsource    = {dblp computer science bibliography, https://dblp.org}
}

@article{chen2018femosaa,
  title={FEMOSAA: Feature-guided and knee-driven multi-objective optimization for self-adaptive software},
  author={Chen, Tao and Li, Ke and Bahsoon, Rami and Yao, Xin},
  journal={ACM Transactions on Software Engineering and Methodology (TOSEM)},
  volume={27},
  number={2},
  pages={1--50},
  year={2018},
  publisher={ACM New York, NY, USA}
}

@article{chen2018sampling,
  author       = {Jianfeng Chen and
                  Vivek Nair and
                  Rahul Krishna and
                  Tim Menzies},
  title        = {"Sampling" as a Baseline Optimizer for Search-Based Software Engineering},
  journal      = {{IEEE} Trans. Software Eng.},
  volume       = {45},
  number       = {6},
  pages        = {597--614},
  year         = {2019},
  url          = {https://doi.org/10.1109/TSE.2018.2790925},
  doi          = {10.1109/TSE.2018.2790925},
  bibsource    = {dblp computer science bibliography, https://dblp.org}
}

@inproceedings{li2021mfes,
  author       = {Yang Li and
                  Yu Shen and
                  Jiawei Jiang and
                  Jinyang Gao and
                  Ce Zhang and
                  Bin Cui},
  title        = {{MFES-HB:} Efficient Hyperband with Multi-Fidelity Quality Measurements},
  booktitle    = {Thirty-Fifth {AAAI} Conference on Artificial Intelligence, {AAAI}
                  2021, Thirty-Third Conference on Innovative Applications of Artificial
                  Intelligence, {IAAI} 2021, The Eleventh Symposium on Educational Advances
                  in Artificial Intelligence, {EAAI} 2021, Virtual Event, February 2-9,
                  2021},
  pages        = {8491--8500},
  publisher    = {{AAAI} Press},
  year         = {2021},
  url          = {https://doi.org/10.1609/aaai.v35i10.17031},
  doi          = {10.1609/AAAI.V35I10.17031},
  bibsource    = {dblp computer science bibliography, https://dblp.org}
}

@inproceedings{xu2015hey,
  author       = {Tianyin Xu and
                  Long Jin and
                  Xuepeng Fan and
                  Yuanyuan Zhou and
                  Shankar Pasupathy and
                  Rukma Talwadker},
  editor       = {Elisabetta Di Nitto and
                  Mark Harman and
                  Patrick Heymans},
  title        = {Hey, you have given me too many knobs!: understanding and dealing
                  with over-designed configuration in system software},
  booktitle    = {Proceedings of the 2015 10th Joint Meeting on Foundations of Software
                  Engineering, {ESEC/FSE} 2015, Bergamo, Italy, August 30 - September
                  4, 2015},
  pages        = {307--319},
  publisher    = {{ACM}},
  year         = {2015},
  url          = {https://doi.org/10.1145/2786805.2786852},
  doi          = {10.1145/2786805.2786852},
  bibsource    = {dblp computer science bibliography, https://dblp.org}
}

@article{chen2024adapting,
  author       = {Tao Chen and
                  Miqing Li},
  title        = {Adapting Multi-objectivized Software Configuration Tuning},
  journal      = {Proc. {ACM} Softw. Eng.},
  volume       = {1},
  number       = {{FSE}},
  pages        = {539--561},
  year         = {2024},
  url          = {https://doi.org/10.1145/3643751},
  doi          = {10.1145/3643751},
  bibsource    = {dblp computer science bibliography, https://dblp.org}
}

@book{cox2000multidimensional,
  title={Multidimensional scaling},
  author={Cox, Trevor F and Cox, Michael AA},
  year={2000},
  publisher={CRC press}
}

@inproceedings{jamshidi2016uncertainty,
  author       = {Pooyan Jamshidi and
                  Giuliano Casale},
  title        = {An Uncertainty-Aware Approach to Optimal Configuration of Stream Processing
                  Systems},
  booktitle    = {24th {IEEE} International Symposium on Modeling, Analysis and Simulation
                  of Computer and Telecommunication Systems, {MASCOTS} 2016, London,
                  United Kingdom, September 19-21, 2016},
  pages        = {39--48},
  publisher    = {{IEEE} Computer Society},
  year         = {2016},
  url          = {https://doi.org/10.1109/MASCOTS.2016.17},
  doi          = {10.1109/MASCOTS.2016.17},
  bibsource    = {dblp computer science bibliography, https://dblp.org}
}

@inproceedings{he2022multi,
  author       = {Haochen He and
                  Zhouyang Jia and
                  Shanshan Li and
                  Yue Yu and
                  Chenglong Zhou and
                  Qing Liao and
                  Ji Wang and
                  Xiangke Liao},
  title        = {Multi-Intention-Aware Configuration Selection for Performance Tuning},
  booktitle    = {44th {IEEE/ACM} 44th International Conference on Software Engineering,
                  {ICSE} 2022, Pittsburgh, PA, USA, May 25-27, 2022},
  pages        = {1431--1442},
  publisher    = {{ACM}},
  year         = {2022},
  url          = {https://doi.org/10.1145/3510003.3510094},
  doi          = {10.1145/3510003.3510094},
  bibsource    = {dblp computer science bibliography, https://dblp.org}
}

@article{DBLP:journals/pvldb/ZhangCLWTLC22,
  author       = {Xinyi Zhang and
                  Zhuo Chang and
                  Yang Li and
                  Hong Wu and
                  Jian Tan and
                  Feifei Li and
                  Bin Cui},
  title        = {Facilitating Database Tuning with Hyper-Parameter Optimization: {A}
                  Comprehensive Experimental Evaluation},
  journal      = {Proc. {VLDB} Endow.},
  volume       = {15},
  number       = {9},
  pages        = {1808--1821},
  year         = {2022},
  url          = {https://www.vldb.org/pvldb/vol15/p1808-cui.pdf},
  doi          = {10.14778/3538598.3538604},
  bibsource    = {dblp computer science bibliography, https://dblp.org}
}

@inproceedings{DBLP:journals/corr/abs-2501-00840,
  author       = {Yulong Ye and
                  Tao Chen and
                  Miqing Li},
  title        = {Distilled Lifelong Self-Adaptation for Configurable Systems},
  booktitle    = {47th {IEEE/ACM} International Conference on Software Engineering,
                  {ICSE} 2025, Ottawa, ON, Canada, April 26 - May 6, 2025},
  pages        = {1333--1345},
  publisher    = {{IEEE}},
  year         = {2025},
  url          = {https://doi.org/10.1109/ICSE55347.2025.00094},
  doi          = {10.1109/ICSE55347.2025.00094},
  bibsource    = {dblp computer science bibliography, https://dblp.org}
}

@inproceedings{chen2021multi,
  author       = {Tao Chen and
                  Miqing Li},
  editor       = {Diomidis Spinellis and
                  Georgios Gousios and
                  Marsha Chechik and
                  Massimiliano Di Penta},
  title        = {Multi-objectivizing software configuration tuning},
  booktitle    = {{ESEC/FSE} '21: 29th {ACM} Joint European Software Engineering Conference
                  and Symposium on the Foundations of Software Engineering, Athens,
                  Greece, August 23-28, 2021},
  pages        = {453--465},
  publisher    = {{ACM}},
  year         = {2021},
  url          = {https://doi.org/10.1145/3468264.3468555},
  doi          = {10.1145/3468264.3468555},
  bibsource    = {dblp computer science bibliography, https://dblp.org}
}

@article{chen2024mmo,
  author       = {Pengzhou Chen and
                  Tao Chen and
                  Miqing Li},
  title        = {{MMO:} Meta Multi-Objectivization for Software Configuration Tuning},
  journal      = {{IEEE} Trans. Software Eng.},
  volume       = {50},
  number       = {6},
  pages        = {1478--1504},
  year         = {2024},
  url          = {https://doi.org/10.1109/TSE.2024.3388910},
  doi          = {10.1109/TSE.2024.3388910},
  bibsource    = {dblp computer science bibliography, https://dblp.org}
}

@article{chen2025accuracy,
  author       = {Pengzhou Chen and
                  Jingzhi Gong and
                  Tao Chen},
  title        = {Accuracy Can Lie: On the Impact of Surrogate Model in Configuration
                  Tuning},
  journal      = {{IEEE} Trans. Software Eng.},
  volume       = {51},
  number       = {2},
  pages        = {548--580},
  year         = {2025},
  url          = {https://doi.org/10.1109/TSE.2025.3525955},
  doi          = {10.1109/TSE.2025.3525955},
  bibsource    = {dblp computer science bibliography, https://dblp.org}
}

@article{back1993overview,
  author       = {Thomas B{\"{a}}ck and
                  Hans{-}Paul Schwefel},
  title        = {An Overview of Evolutionary Algorithms for Parameter Optimization},
  journal      = {Evol. Comput.},
  volume       = {1},
  number       = {1},
  pages        = {1--23},
  year         = {1993},
  url          = {https://doi.org/10.1162/evco.1993.1.1.1},
  doi          = {10.1162/EVCO.1993.1.1.1},
  bibsource    = {dblp computer science bibliography, https://dblp.org}
}

@inproceedings{zhang2021restune,
  author       = {Xinyi Zhang and
                  Hong Wu and
                  Zhuo Chang and
                  Shuowei Jin and
                  Jian Tan and
                  Feifei Li and
                  Tieying Zhang and
                  Bin Cui},
  editor       = {Guoliang Li and
                  Zhanhuai Li and
                  Stratos Idreos and
                  Divesh Srivastava},
  title        = {ResTune: Resource Oriented Tuning Boosted by Meta-Learning for Cloud
                  Databases},
  booktitle    = {{SIGMOD} '21: International Conference on Management of Data, Virtual
                  Event, China, June 20-25, 2021},
  pages        = {2102--2114},
  publisher    = {{ACM}},
  year         = {2021},
  url          = {https://doi.org/10.1145/3448016.3457291},
  doi          = {10.1145/3448016.3457291},
  bibsource    = {dblp computer science bibliography, https://dblp.org}
}

@inproceedings{ma2025faster,
author = {Ma, Youpeng and Chen, Tao and Li, Ke},
title = {Faster Configuration Performance Bug Testing with Neural Dual-Level Prioritization},
year = {2025},
publisher = {IEEE Press},
url = {https://doi.org/10.1109/ICSE55347.2025.00201},
booktitle = {Proceedings of the IEEE/ACM 47th International Conference on Software Engineering},
pages = {988–1000},
numpages = {13}
}

@article{nair2018finding,
  author       = {Vivek Nair and
                  Zhe Yu and
                  Tim Menzies and
                  Norbert Siegmund and
                  Sven Apel},
  title        = {Finding Faster Configurations Using {FLASH}},
  journal      = {{IEEE} Trans. Software Eng.},
  volume       = {46},
  number       = {7},
  pages        = {794--811},
  year         = {2020},
  url          = {https://doi.org/10.1109/TSE.2018.2870895},
  doi          = {10.1109/TSE.2018.2870895},
  bibsource    = {dblp computer science bibliography, https://dblp.org}
}

@inproceedings{han2016empirical,
  author       = {Xue Han and
                  Tingting Yu},
  title        = {An Empirical Study on Performance Bugs for Highly Configurable Software
                  Systems},
  booktitle    = {Proceedings of the 10th {ACM/IEEE} International Symposium on Empirical
                  Software Engineering and Measurement, {ESEM} 2016, Ciudad Real, Spain,
                  September 8-9, 2016},
  pages        = {23:1--23:10},
  publisher    = {{ACM}},
  year         = {2016},
  url          = {https://doi.org/10.1145/2961111.2962602},
  doi          = {10.1145/2961111.2962602},
  bibsource    = {dblp computer science bibliography, https://dblp.org}
}

@article{chen2018survey,
  author       = {Tao Chen and
                  Rami Bahsoon and
                  Xin Yao},
  title        = {A Survey and Taxonomy of Self-Aware and Self-Adaptive Cloud Autoscaling
                  Systems},
  journal      = {{ACM} Comput. Surv.},
  volume       = {51},
  number       = {3},
  pages        = {61:1--61:40},
  year         = {2018},
  url          = {https://doi.org/10.1145/3190507},
  doi          = {10.1145/3190507},
  bibsource    = {dblp computer science bibliography, https://dblp.org}
}

@article{chen2015toward,
  author       = {Tao Chen and
                  Rami Bahsoon},
  title        = {Toward a Smarter Cloud: Self-Aware Autoscaling of Cloud Configurations
                  and Resources},
  journal      = {Computer},
  volume       = {48},
  number       = {9},
  pages        = {93--96},
  year         = {2015},
  url          = {https://doi.org/10.1109/MC.2015.278},
  doi          = {10.1109/MC.2015.278},
  bibsource    = {dblp computer science bibliography, https://dblp.org}
}

\end{document}